\documentclass[a4paper,onecolumn,12pt]{article}
\usepackage{amsmath}
\DeclareMathOperator*{\argmax}{arg\,max}
\DeclareMathOperator*{\argmin}{arg\,min}
\usepackage{amsfonts}
\usepackage{xcolor}
\definecolor{winered}{rgb}{0.5,0,0}
\usepackage{float}
\usepackage{placeins}
\usepackage[cal=boondoxo]{mathalfa}
\usepackage{tikz}
\usetikzlibrary{fit,positioning}
\usepackage{lscape}
\usepackage{longtable}
\usepackage{epstopdf}
\usepackage{float}
\usepackage{setspace}
\usepackage{textcomp}
\usepackage{natbib}

\usepackage{algorithm}
\usepackage[end]{algpseudocode}
\algnewcommand\algorithmicto{\textbf{to }}
\usepackage{etoolbox}
\patchcmd{\thebibliography}{\section*{\refname}}{}{}{}

\newcommand{\bm}[1]{\boldsymbol{#1}}

\usepackage[bookmarks=true, bookmarksnumbered=true, allbordercolors={1 1 1}]{hyperref}
\hypersetup{
  colorlinks   = true, 
  urlcolor     = blue, 
  linkcolor    = blue, 
  citecolor    = winered,
}

\usepackage[nameinlink,capitalise]{cleveref}

\begin{document}

\title{{\LARGE Bayesian dynamic variable selection in high dimensions}}
\author{Gary Koop \\
\emph{University of Strathclyde} \and Dimitris Korobilis \thanks{Corresponding Author: Adam Smith Business School, University of Glasgow, G12 8QQ Glasgow, UK, email: \href{mailto:Dimitris.Korobilis@glasgow.ac.uk}{Dimitris.Korobilis@glasgow.ac.uk}} \\
\emph{University of Glasgow}}
\date{ }
\maketitle

\begin{abstract}
\noindent This paper proposes a variational Bayes algorithm for computationally efficient posterior and predictive inference in time-varying parameter (TVP) models. Within this context we specify a new dynamic variable/model selection strategy for TVP dynamic regression models in the presence of a large number of predictors. This strategy allows for assessing in individual time periods which predictors are relevant (or not) for forecasting the dependent variable. The new algorithm is evaluated numerically using synthetic data and its computational advantages are established. Using macroeconomic data for the US we find that regression models that combine time-varying parameters with the information in many predictors have the potential to improve forecasts of price inflation over a number of alternative forecasting models.
\thispagestyle{empty} 

\bigskip \bigskip

\noindent  \emph{Keywords:} dynamic linear model; approximate posterior inference; dynamic variable selection; forecasting

\bigskip \medskip
\noindent \emph{JEL Classification:}\ C11, C13, C52, C53, C61
\end{abstract}

\newpage
\onehalfspacing
\setcounter{page}{1}

\section{Introduction}
Regression models that incorporate stochastic variation in parameters have been used by economists at least since the work \cite{CooleyPrescott1976}. Thirty years later, \cite{Granger2008} argued that time-varying parameter models might become the norm in econometric inference since, as he illustrated via White's theorem, time variation is able to approximate generic forms of nonlinearity in parameters. Indeed, initiated by the unprecedented shocks observed during and after the Global Recession of 2007-9, a large recent literature has established the importance of modeling time variation in the intercept, slopes and variance of regressions for forecasting economic time series; see \cite{StockWatson2007} for a representative example of a model using only a stochastic intercept and volatilities. At the same time, the stylized fact that economic predictors are short-lived -- that is, relevant for the dependent variable only in short periods\footnote{An alternative terminology for such periods, which is due to \cite{Farmeretal2018}, is ``pockets of predictability''.} -- has emerged in various forecasting problems such as inflation \citep{KoopKorobilis2012}, stock returns \citep{DanglHalling2012} and exchange rates \citep{Byrneetal2018}. Following these observations, there is no shortage of recent econometric work on methods for penalized estimation of time-varying parameter models via classical or Bayesian shrinkage, as well as variable selection methods; see for example \cite{Belmonteetal2014}, \cite{Bitto2019}, \cite{KalliGriffin2014}, \cite{CallotKristensen2014}, \cite{Korobilis2019}, \cite{Kowaletal2019}, \cite{NakajimaWest2013}, \cite{RockovaMcAlinn2017}, \cite{UribeLopes2017} and \cite{YousufNg2019}.

In this paper we add to this literature by proposing a new dynamic variable selection prior and a novel, for the field of economics, Bayesian estimation methodology. In particular, we propose to use variational Bayes (VB) inference to estimate time-varying parameter regressions using state-space methods. Variational inference has long been used in data science problems such as large-scale document analysis, computational neuroscience, and computer vision \citep{Bleietal2017}. Nevertheless, it is only relatively recently that posterior consistency and other theoretical properties of these methods have been explored by mainstream statisticians \citep{WangBlei2019}. Variational inference is a unified estimation methodology which shares similarities with the Gibbs sampler that many economists traditionally use to estimate time-varying parameter models \citep[see for example][]{StockWatson2007}. Like the Gibbs sampler, parameter updates are derived for one parameter at a time conditional on all other parameters using an iterative scheme. Unlike the Gibbs sampler, there is no repeated sampling involved and the output of VB is typically the first two moments of the posterior distribution of parameters. Our first task is to introduce this estimation scheme in the context of TVP regressions, and contrast it to existing estimation algorithms used in economics for capturing structural change.

Our second contribution lies on the development of a dynamic variable selection prior that is a conceptually straightforward extension of the static variable selection prior of \cite{GeorgeMcCulloch1993}. The dynamic extension of this prior allows to tackle the non-trivial econometric problem of allowing some predictor variables to enter the TVP regression, model only in some periods of the full estimation sample. With $p$ predictors and $T$ time periods, dynamic variable selection involves choosing the ``best'' among $2^{p}$ models at each point in time $t$, for $t=1,...,T$. Such procedure is in line with strong, recent empirical evidence that different factors might be driving predictability of economic variables over time; see \cite{Rossi2013} for a thorough review of this idea. By specifying our new prior within a variational Bayes framework, we are able to derive an algorithm that is numerically stable and can be extended to much larger $p$ and $T$ than was possible before.\footnote{In particular, many of the algorithms cited above, such as \cite{KoopKorobilis2012}, \cite{KalliGriffin2014}, or \cite{NakajimaWest2013}, are unable to scale up to regressions with hundreds of predictors.}

We show, via a Monte Carlo exercise and an empirical application, that our proposed algorithm works well in high-dimensional, sparse, time-varying parameter settings. Using artificial data we establish that the new algorithm is precise in estimation and in dynamic variable selection, even in settings with more predictors than time-series observations. In a forecasting exercise of various measures of price inflation, we illustrate that our methodology applied to a time-varying parameter regression with 400+ predictors is able to beat a wide range of linear and nonlinear forecasting regressions. The empirical results provide strong evidence that the new algorithm can achieve estimation accuracy comparable to Markov chain Monte Carlo algorithms, while being much faster to run. The additional feature of dynamic variable selection successfully prevents overparametrization, since our high-dimensional TVP specification is able to beat both parsimonious time series models with no predictors as well as factor models and penalized likelihood estimators.

The remainder of the paper proceeds as follows. Section 2 introduces the basic principles of VB inference for approximating intractable posteriors, and applies these principles to the problem of estimating a simplified time-varying parameter regression model. Section 3 introduces the the novel modelling assumptions, namely dynamic variable selection and stochastic volatility, and derives an estimation algorithm within the VB framework. Section 4 assesses the new algorithm on simulated data. In Section 5 we apply the new methodology to the problem of forecasting US inflation using time-varying parameter regressions with many predictors.

\section{Variational Bayes inference in state-space models}
as variational Bayes (VB) is not an established estimation methodology in econometrics, we first provide a generic discussion of VB methods in approximating intractable posterior distributions. We then apply the generic concepts and formulas to the specific problem of estimating a simplified time-varying parameter regression model with known measurement error variance.\footnote{Readers already familiar with these concepts can skim through this section, and focus on our novel methodology that is described in the following section} Detailed reviews of variational Bayes can be found in \cite{Bleietal2017} and \cite{OrmerodWand2010}, among several others. Variational Bayes estimation of state-space models is described in detail in the monograph of \cite{smidl2006variational}, as well as research papers such as \cite{BealGhahramani2003}, \cite{Tranetal2017}, and \cite{Wangetal2016}.

\subsection{Basics of variational Bayes}
Consider data $y$, latent variables $s$ and (latent) parameters $\theta$. Our interest lies in time-varying parameter models that admit a state-space form. Hence, $s$ represents unobserved state variables, such as time-varying regression coefficients and time-varying measurement error variances, and $\theta$ represents all other parameters, such as the error covariances in the state equation. The joint posterior of interest is $p\left(s, \theta |y\right)$ with associated marginal likelihood $p\left( y\right) $ and joint density of data and parameters $p\left(y,s,\theta \right)$. When the joint posterior is complex and computationally intractable, we can define an approximating density $q\left( s,\theta \vert y \right)$ that belongs to a family $\mathcal{F}$ of simpler distributions defined over the parameter space spanned by $s,\theta$. The main idea behind variational Bayes inference is to make this approximating posterior distribution $q\left( s,\theta \vert y \right)$ as close as possible to $p\left( s,\theta |y\right) $, where ``distance'' is measured with the Kullback-Leibler divergence\footnote{For notational simplicity we henceforth abbreviate multiple integrals using a single integration symbol.}
\begin{equation}
KL\left(q \vert \vert p \right)=\int q\left( s, \theta \vert y \right) \log \left\{ \frac{q\left( s, \theta \vert y \right) }{p\left( s, \theta | y \right) }\right\} \mathrm{d} s \mathrm{d} \theta .  \label{KL}
\end{equation}%
That is, the aim is to find the optimal $q^{\star}\left( s,\theta \vert y \right)$ that solves
\begin{equation}
q^{\star}\left( s,\theta \vert y \right) = \argmin_{q\left( s,\theta \vert y \right) \in \mathcal{F}} KL\left(q \vert \vert p \right).
\end{equation}
Insight for why $KL\left(q \vert \vert p \right)$ is a desirable distance metric arises from a simple re-arrangement involving the log of the marginal likelihood \citep[page 142]{OrmerodWand2010} where it can be shown that
\begin{eqnarray}
\log p\left( y\right) & = &\log p\left( y\right) \int p\left(s,\theta \vert y \right)\mathrm{d} s \mathrm{d} \theta = \int p\left(s,\theta \vert y \right) \log p\left( y\right)\mathrm{d} s \mathrm{d} \theta \\
& = &  \int q(s, \theta \vert y) \log \left \lbrace \frac{p\left(y,s,\theta \right) / q(s, \theta \vert y) }{p(s,\theta \vert y)/q(s, \theta \vert y)}  \right \rbrace \mathrm{d} s \mathrm{d} \theta \\ & = &
\int q\left( s, \theta \vert y \right) \log \left\{ \frac{%
p\left( y,s,\theta \right) }{q\left( s,\theta \vert y \right) }\right\} \mathrm{d} s \mathrm{d} \theta + KL\left(q \vert \vert p \right). \label{log_marg_lik}
\end{eqnarray}
Because $KL\left(q \vert \vert p \right)$ is non-negative (it is exactly zero when $q\left( s,\theta \vert y \right) = p\left( s,\theta |y\right) $), the quantity
{\footnotesize
\begin{equation}
\mathcal{G} \left(q(s,\theta \vert y )\right)  = \exp\left[ \int q\left( s,\theta \vert y \right) \log \left\{ 
\frac{p\left( y, s, \theta \right) }{q\left( s, \theta \vert y \right) }\right\} d s d\theta \right] %
 \equiv \exp \left [ \mathbb{E}_{q\left( s,\theta \vert y \right)}\left( \log\left( p\left( y, s, \theta \right)\right) - \log \left( q\left( s, \theta \vert y \right) \right) \right) \right],
\end{equation}
}becomes a lower bound for the marginal likelihood $ p(y)$.\footnote{In the following we denote as $\mathbb{E}_{q(\bullet)}$ the expectation w.r.t to a function $q(\bullet)$.} The function $ \mathcal{G} \left(q(s,\theta \vert y )\right)$ is known as the Evidence Lower Bound (ELBO). Therefore, instead of minimizing the objective function $KL\left(q \vert \vert p \right)$ (which cannot be evaluated) we can find an approximating density $q^{\star}\left( s,\theta \vert y \right)$ that maximizes the marginal data density $p\left( y \right)$ by maximizing the ELBO. We emphasize that $\mathcal{G}$ is a functional on the distribution $q(s,\theta \vert y)$. As a result, the ELBO can be maximized iteratively using calculus of variations.

If we assume for simplicity the so-called (in Physics) \emph{mean field} factorization of the form $q\left( s, \theta \vert y \right)=q\left( \theta \vert y \right) q\left( s \vert y \right)$, it can be shown\footnote{A formal and thorough derivation of these ideas is given in the excellent monograph of \cite{smidl2006variational}; see Theorem 3.1 and subsequent results.} that the optimal choices for $q\left( s \vert y\right)$ and $q\left( \theta \vert y \right)$ are
\begin{eqnarray}
q\left( s \vert y \right) & \propto & \exp \left[ \int q(\theta \vert y) \log p\left(s \vert y,\theta \right) \mathrm{d} \theta \right] \equiv  \exp \left[ \mathbb{E}_{q(\theta \vert y)}\left( \log p\left(s \vert y, \theta \right) \right) \right], \label{VBEM1} \\
q\left( \theta \vert y \right) & \propto & \exp \left[  \int q(s \vert y)  \log p\left(\theta \vert y,s \right)  \mathrm{d} s  \right] \equiv \exp \left[ \mathbb{E}_{q(s \vert y)}\left( \log p\left(\theta \vert y, s \right) \right) \right]. \label{VBEM2}
\end{eqnarray} 
The first expression denotes the expectation over $q(\theta \vert y)$ of the conditional posterior for $s$, and the second expression denotes the expectation over $q(s \vert y)$ of the conditional posterior for $\theta$. Because $q(\theta \vert y)$ is a function of $q(s \vert y)$, and vice-versa, the above quantities can be approximated iteratively instead of relying on more computationally expensive numerical optimization techniques. Given an initial guess regarding the values of $(\theta, s)$, VB algorithms iterate over these two quantities until $\mathcal{G} \left(q(s,\theta \vert y)\right)$ has reached a maximum. Due to similarities with the Expectation-Maximization (EM) algorithm of \cite{Dempsteretal1977}, this iterative procedure in its general form is sometimes referred to as the \emph{Variational Bayesian EM (VB-EM)} algorithm; see \cite{BealGhahramani2003}. It is also worth noting the relationship with Gibbs sampling. Like Gibbs sampling, equations \eqref{VBEM1} and \eqref{VBEM2} involve the full conditional posterior distributions. But unlike Gibbs sampling, the VB-EM algorithm does not repeatedly simulate from them and is computationally much faster.

\subsection{VB estimation of a simple TVP regression model} \label{sec:simple_TVP}
Before collecting all building blocks of our proposed methodology, we outline a VB algorithm for the univariate TVP regression with known measurement error variance $\underline{\sigma}^{2}$. This simplified model is of the form
\begin{eqnarray}
y_{t} &=& \mathbf{x}_{t} \bm{\beta}_{t}+ \varepsilon _{t}  \label{TVP_measurement} \\
\bm{\beta}_{t} &=& \bm{\beta}_{t-1} + \bm \eta_{t}  \label{TVP_state}
\end{eqnarray}%
where $y_{t}$ is the time $t$ scalar value of the dependent variable, $t=1,..,T$, $\mathbf{x}_{t}$ is a $1 \times p$ vector of exogenous predictors and lagged dependent variables, $\varepsilon _{t} \sim N\left( 0,\underline{\sigma}^{2} \right) $, $\bm \eta_{t} \sim N\left( 0,\bm W_{t} \right)$ with $ \bm{W}_{t}=diag\left(w_{1,t},...,w_{p,t} \right)$ a $p \times p$ diagonal matrix\footnote{By restricting $\bm W_t$ not to be a full covariance matrix, coefficients $\beta_{it}$ and $\beta_{jt}$ are uncorrelated a-posteriori for $i \neq j$, which might not seem like an empirically relevant assumption. However, allowing for cross-correlation in the state vector $\bm \beta_{t}$ can result in counterproductive increases in estimation uncertainty, with this problem being significantly more pronounced in higher dimensions. A diagonal $\bm W_{t}$ allows for a more parsimonious econometric specification, less cumbersome derivations of posterior distributions, and faster and numerically stable computation; see also \cite{Belmonteetal2014}, \cite{Bitto2019} and \cite{RockovaMcAlinn2017} who adopt a similar assumption.} and $\bm w_{t} = [w_{1,t},...,w_{p,t}]^{\prime}$ a $p \times 1$ vector. In likelihood-based analysis of state-space models it simplifies inference if it is assumed that $\varepsilon _{t}$ and $\bm \eta_{t}$ are independent of one another and we do adopt this assumption here. Finally, we use a notational convention where $j,t$ subscripts denote the $j^{th}$ element of a time varying state variable, or parameter, observed only at time $t$, while $1:t$ subscripts denote all the observations of a state variable from period $1$ up to period $t$.

The model in equations \eqref{TVP_measurement} and \eqref{TVP_state} has unknown parameters $\left( \bm \beta_{1:T},\bm w_{1:T} \right)$. Following the analysis of the previous subsection we first consider the independent prior on the initial conditions $\bm\beta_{0},\bm w_{0}$ of the form
\begin{equation}
p(\bm \beta_{0},\bm w_{0})  =  p(\bm \beta_{0}) \prod_{j=1}^{p} p(w_{j,0}) =  N(\bm m_{0},\bm P_{0}) \times \prod_{j=1}^{p} \left[ Gamma( c_{j,0}, d_{j,0}) \right]^{-1}, \label{ssm_prior}
\end{equation}
where $w_{j,0}$ is the $j^{th}$ element of $w_{0}$, and $Gamma(a,b)$ denotes the Gamma distribution with shape parameter $a$ and rate parameter $b$, that is, the definition of the Gamma distribution that has mean $a/b$ and variance $a/b^{2}$. The time $t$ prior, conditional on observing information up to time $t-1$, is given by the Chapman-Kolmogorov equation
\begin{equation}
\begin{array}{lll}
p(\bm \beta_{t},\bm w_{t} \vert \bm y_{1:t-1}) & = &  \int_{\mathcal{B},\mathcal{W}} p(\bm \beta_{t} \vert \bm \beta_{t-1}) \prod_{j=1}^{p} p(w_{j,t} \vert w_{j,t-1}) \\
& \times & p(\bm \beta_{t-1}, \bm w_{t-1} \vert y_{t-1}) d \bm\beta_{t-1} dw_{1,t-1}...dw_{p,t-1},
\end{array} \label{detailed_decomp}
\end{equation}
where $\mathcal{B}$ is the support of $\bm \beta_{t}$ and $\mathcal{W}$ the support of all $w_{j,t}$. Finally, once the measurement $y_{t}$ is observed, we obtain from Bayes theorem the following time $t$ posterior distribution
\begin{equation}
p(\bm \beta_{t},\bm w_{t} \vert \bm y_{1:t}) \propto p(y_{t} \vert \bm \beta_{t},\bm w_{t}) p(\bm \beta_{t},\bm w_{t} \vert \bm y_{1:t-1}).
\end{equation}

This Bayesian joint posterior distribution is rarely analytically tractable, even if conjugate prior densities have been specified. However, posterior conditionals can be tractable, and this is why in macroeconomics TVP models are predominantly estimated using the Gibbs sampler; see \cite{StockWatson2007} for an example. Nevertheless, sampling repeatedly using (Markov chain) Monte Carlo methods is computationally prohibitive in high-dimensional settings or in settings with more flexible likelihood and prior distributions. For that reason we define a tractable variational density as an approximation to the exact intractable time $t$ posterior, that is, we define $p\left(\bm \beta_{t},\bm w_{t} \vert \bm y_{1:t} \right)  \approx  q\left( \bm \beta_{t}, \bm w_{t} \vert \bm y_{1:t} \right)$. Among all possible functions $q\left( \bm \beta_{t}, \bm w_{t} \vert \bm y_{1:t} \right)$ we want to obtain the one that has hyperparameters that minimize the relative entropy with the true posterior. Following the discussion earlier in this section, this problem is equivalent to maximizing the evidence lower bound (ELBO) of the log-marginal likelihood, that is, it is the solution to
{\small
\begin{equation}
q^{\star}\left( \bm \beta_{t} , \bm w_{t} \vert \bm y_{1:t} \right)  = 
\argmax_{q\left( \bm \beta_{t}, \bm w_{t} \vert \bm y_{1:t} \right)} \int q\left( \bm \beta_{t}, \bm w_{t} \vert \bm y_{1:t} \right) \log \left(\frac{q\left( \bm \beta_{t} , \bm w_{t} \vert \bm y_{1:t} \right)}{p\left(\bm \beta_{t}, \bm w_{t} \vert \bm y_{1:t} \right)} \right). 
\label{KL_detailed}
\end{equation} }

This maximization problem is simplified once we assume the mean field factorization of the form $q \left( \bm \beta_{t} , \bm w_{t} \vert \bm y_{1:t} \right) = q\left( \bm \beta_{t} \vert \bm y_{1:t} \right) \prod q\left(w_{j,t} \vert \bm y_{1:t} \right)$ so we can optimize $\bm \beta_{t}$ and $\bm w_{t}$ sequentially. As a result, using variational calculus \citep{smidl2006variational} we can show that the ELBO is maximized by iterating through the following recursions
\begin{eqnarray}
 q\left( \bm \beta_{t} \vert \bm y_{1:t} \right)  & \propto &  \exp \left( \int \log p\left(y_{t},\bm \beta_{t}, \bm w_{t} \vert \bm y_{1:t-1} \right) \prod_{j} q\left(w_{j,t} \vert \bm y_{1:t} \right) d \bm w_{t} \right), \label{variational_calculus1}\\
 q\left( w_{j,t} \vert \bm y_{1:t} \right)  & \propto & \exp \left( \int \log p\left(y_{t},\bm \beta_{t}, \bm w_{t} \vert \bm y_{1:t-1} \right) q\left(\bm \beta_{t} \vert \bm  y_{1:t} \right) d \bm \beta_{t} \right), j=1,...,p.  \label{variational_calculus2}
\end{eqnarray}
Both formulas above become equalities after the addition of a normalizing constant. The first expression is an expectation with respect to the probability density $\prod_{j} q\left(w_{j,t} \vert \bm y_{1:t-1} \right)$, that is, we can write equation \eqref{variational_calculus1} using the following form
\begin{eqnarray}
q\left( \bm \beta_{t} \vert \bm y_{1:t} \right) & \propto & \exp \left( \mathbb{E}_{q\left(\bm w_{t} \vert \bm y_{1:t} \right)} \left( \log p\left(y_{t},\bm \beta_{t}, \bm w_{t} \vert \bm y_{1:t-1} \right) \right)  \right) \label{first_part}\\
&  = & \exp \left( \mathbb{E}_{q\left(\bm w_{t} \vert \bm y_{1:t} \right)} \left( \log \left[ p\left(y_{t} \vert \bm \beta_{t}, \bm w_{t} \right)p \left( \bm \beta_{t}\vert \bm y_{1:t-1} \right)p \left( \bm w_{t} \vert \bm y_{1:t-1} \right) \right] \right) \right) \label{second_part} \\
& = & p\left(y_{t} \vert \bm \beta_{t}, \bm w_{t} \right)\exp \left( \mathbb{E}_{q\left(\bm w_{t} \vert \bm y_{1:t} \right)} \left( \log  p \left( \bm \beta_{t}\vert \bm \beta_{t-1} \right) + \log q\left(\bm \beta_{t} \vert \bm y_{1:t-1}\right)  \right) \right) \label{third_part} 
\end{eqnarray}
where $q\left(\bm \beta_{t} \vert y_{1:t-1}\right)$ is the time $t$ prior of $\bm \beta_{t}$ obtained from the time $t-1$ posterior $q\left( \bm \beta_{t-1} \vert \bm y_{1:t-1}\right)$ using the Kalman filter recursions, and the term $p \left( \bm  w_{t} \vert \bm y_{1:t-1} \right)$ in \eqref{second_part} disappears because the expectation is w.r.t the variational posterior of $w_{t}$. This latter representation of $q\left( \bm  \beta_{t} \vert \bm y_{1:t} \right)$ can be trivially updated by a Normal distribution, with moments given by the Kalman filter and smoother; see \citet[Chapter 7]{smidl2006variational} for detailed derivations. We can use similar arguments in order to show that equation \eqref{variational_calculus2} is an expectation that leads to a $q\left( \bm w_{t}^{-1} \vert \bm y_{1:t} \right)$ of the form $G \left(c_{j,t},d_{j,t}\right)$ (or equivalently to $q\left( \bm w_{t} \vert \bm y_{1:t} \right)$ that is inverse Gamma). 

\begin{algorithm}[H]
\caption{\textit{Variational Bayes algorithm for TVP regression model with fixed measurement variance}}\label{algorithm:VBKF_offline}
  \scriptsize
\begin{algorithmic}[1]
\State Choose values of hyperparameters $\bm m_{0},\bm P_{0},c_{j,0},d_{j,0}$ for $j=1,...,p$. Set $\mathcal{r}=1$ and initialize $W^{(\mathcal{r}-1)}$. 
\While{$ \Vert \mathcal{G}\left(q(\bm \beta^{(\mathcal{r})},\bm w^{(\mathcal{r})} \vert \bm y \right) - \mathcal{G}\left(q(\bm \beta^{(\mathcal{r}-1)},\bm w^{(\mathcal{r}-1)} \vert \bm y \right) \Vert \rightarrow 0$}
\State \textbf{Step 1:} Approximate, $\forall$ $t=1,...,T$, the posterior $$q^{\mathcal{r}} \left( \bm \beta_{t} \vert \bm  y_{1:T} \right) \sim N \left( \bm m^{\mathcal{r}}_{t},\bm P^{\mathcal{r}}_{t} \right)$$
conditional on $\bm W^{(\mathcal{r}-1)}, \underline{\sigma}^{2}$, where $\bm m_{t}^{\mathcal{r}}, \bm P_{t}^{\mathcal{r}}$ $\forall$ $t=1,...,T$, are obtained using the Kalman filter and the Rauch-Tung-Striebel smoother
\State \textbf{Step 2:} Approximate, $\forall$ $t=1,...,T$ and $j=1,...,p$, the posterior
$$
q^{\mathcal{r}} \left( w_{j,t}^{-1} \vert \bm y_{1:T} \right) \sim G \left(c_{j,t}^{\mathcal{r}},d_{j,t}^{\mathcal{r}} \right)
$$
conditional on $\bm \mu_{t}^{\mathcal{r}}, \bm P_{t}^{\mathcal{r}}$, where $c_{j,t}^{\mathcal{r}} = c_{j,0} + 1/2$, $d_{j,t}^{\mathcal{r}} = d_{j,0} + \bm D_{jj,t}/2$ with $\bm D_{t} = \left( \bm P_{t}^{\mathcal{r}} - \bm P_{t-1}^{\mathcal{r}} \right) + \left( \bm m_{t}^{(\mathcal{r})} \bm m_{t}^{(\mathcal{r}) \prime} - \bm m_{t-1}^{(\mathcal{r})} \bm m_{t-1}^{(\mathcal{r}) \prime} \right)$. Set $\bm W^{(\mathcal{r})} = diag \left( d_{1,t}^{\mathcal{r}}/c_{1,t}^{\mathcal{r}},...,d_{p,t}^{\mathcal{r}}/c_{p,t}^{\mathcal{r}} \right)$.
\State $\mathcal{r} = \mathcal{r} + 1$
\EndWhile
\State Upon convergence set $q^{\star} \left(\bm \beta_{1:T}, \bm w_{1:T} \vert \bm y_{1:T} \right) = q^{\mathcal{r}} \left( \bm \beta_{1:T} \vert \bm y_{1:T} \right) \times \prod_{j=1}^{p} q^{\mathcal{r}} \left( w_{j,t} \vert \bm y_{1:T} \right)$ using the parameters $( \bm m_{1:T}^{\mathcal{r}}, \bm P_{1:T}^{\mathcal{r}},\bm c_{1:p,1:T}^{\mathcal{r}},\bm d_{1:p,1:T}^{\mathcal{r}})$ obtained during the last iteration of the \emph{while} loop.
\end{algorithmic}
\end{algorithm}

Algorithm \autoref{algorithm:VBKF_offline} provides pseudocode for the basic VB estimation problem described in this section, without assuming either a (dynamic) variable selection prior or stochastic volatility in the measurement equation. In the following section we drop these two unrealistic assumptions.

\section{Variational Bayes Inference in High-Dimensional TVP Regressions}
We rewrite for convenience the univariate time-varying parameter model
\begin{eqnarray}
y_{t} &=& \mathbf{x}_{t} \bm{\beta}_{t}+ \varepsilon _{t}  \label{TVP_measurement_new} \\
\bm{\beta}_{t} &=& \bm{\beta}_{t-1} + \bm \eta_{t}  \label{TVP_state_new}
\end{eqnarray}
where we define now $\varepsilon_{t} \sim N(0,\sigma_{t}^{2})$ with $\sigma_{t}^{2}$ a stochastic (time-varying) variance parameter, and we assume that the dimension $p$ of $\bm \beta_{t} = \left(\beta_{1,t},...,\beta_{p,t} \right)^{\prime}$ is large and possibly $p \gg T$.

\subsection{Dynamic variable selection and averaging} \label{sec:dvs}
The core ingredient of our modeling approach is a dynamic variable/model selection strategy. We specify a dynamic variable selection (DVS) prior that extends the ``static'' variable selection prior of \cite{GeorgeMcCulloch1993} that was originally developed for the constant parameter regression using MCMC and is of the form
\begin{eqnarray}
\beta_{j,t} \vert \gamma_{j,t}, \tau_{j,t}^{2} & \sim & \left( 1 - \gamma_{j,t} \right) N \left( 0, \underline{c} \times \tau_{j,t}^{2} \right) + \gamma_{j,t} N \left( 0, \tau_{j,t}^{2} \right), \label{dSSVS1}\\
\gamma_{j,t} \vert \pi_{t} &\sim & Bernoulli \left(\pi_{0,t} \right), \label{dSSVS2} \\
\frac{1}{\tau_{j,t}^{2}} & \sim & Gamma \left(g_{0}, h_{0} \right)  \label{dSSVS3} \\ 
\pi_{0,t} & \sim & Beta(1,1), \label{dSSVS4}
\end{eqnarray}
for $j=1,...,p$, where $\underline{c}$, $g_0$ and $h_{0}$ are fixed prior hyperparameters. Variable selection principles require us to set $\underline{c} \rightarrow 0$, such that the first component in the prior for $\beta_{j,t}$ shrinks the posterior towards zero, while the second component has variance $\tau_{j,t}^{2}$ which is ``large enough'' in order to allow for unrestricted estimation. The choice between the two components in the prior for $\beta_{j,t}$ is governed by the random variable $\gamma_{j,t}$ which is distributed Bernoulli and takes values either zero or one. If $\gamma_{j,t}=1$ the prior for $\beta_{j,t}$ has a Normal prior with zero mean and variance $\tau_{j,t}^{2}$, while if $\gamma_{j,t}=0$ the prior variance becomes $\underline{c}\tau_{j,t}^{2}$. 

Early papers such as \cite{GeorgeMcCulloch1993} give very broad guidelines on choosing values for $\underline{c}$ and $\tau_{j,t}^{2}$ such that the first component in equation \eqref{dSSVS1} has small enough variance (to force shrinkage) and the second component has large enough variance (to allow unrestricted estimation). More recently, \cite{NarisettyHe2014} show that selecting and fixing the prior variances of such mixture priors could, as $T$ and $p$ grow, lead to model selection inconsistency. The authors suggest to specify these parameters to be certain deterministic functions of the data dimensions $T$ and $p$. In our case, we do fix $\underline{c}=10^{-4}$ such that the first component has always smaller variance, but we assume $\left(\tau_{j,t}^{2}\right)^{-1}$ is a random variable that has a Gamma prior. That way this parameter is always updated by the information in the data likelihood. The choice of a Gamma prior for $\left(\tau_{j,t}^{2}\right)^{-1}$ implies that the marginal prior for $\beta_{j,t}$ is a mixture of leptokurtic Student's T distributions whose components could tend to shrink $\beta_{j,t}$ towards zero, regardless of whether $\gamma_{j,t}$ is zero or one. Therefore, the proposed prior is able to find patterns of dynamic sparsity as well as impose dynamic shrinkage in time-varying parameters, a property that is very desirable in high-dimensional settings.\footnote{In signal processing a signal (regression coefficient vector) is typically sparse by default, that is, the researcher knows a-priori to expect that estimates of several coefficients will tend to be exactly zero. In economics, the sparsity assumption might not be empirically founded in certain settings; see the discussion in \cite{Giannoneetal2017}. In such cases, a dense model may be preferred, that is, a model where all predictors are relevant with varying weights. While factor models and principal components have been used widely to model dense models in macroeconomics, shrinkage methods are also quite reliable for this task. In particular, we note the result in \cite{DeMoletal2008} that forecasts from Bayesian shrinkage are highly correlated to forecasts from principal components.} 
 
Finally, it becomes apparent that under this variable selection prior setting, $\widehat{\pi}_{0,t} = \mathbb{E} \left( p\left(\pi_{0,t}\right) \right)=\frac{1}{2}$ is the time $t$ prior mean probability of inclusion of all predictors in the TVP regression, while the quantity $\widetilde{\pi}_{j,t} = \mathbb{E} \left( p\left( \gamma_{j,t} \vert \bm y_{1:T} \right)\right)$ is the posterior mean probability of inclusion in the regression of predictor $j$ at time period $t$, simply referred to as the \emph{posterior inclusion probability (PIP)}. Due to the fact that all of the hyperparameters $\bm \gamma,\pi$ and $\bm \tau^{2}$ are time-varying, our prior allows to obtain time-varying PIPs whose interpretation extends this of PIPs in constant parameter settings, such as the one in \cite{GeorgeMcCulloch1993}, in a straightforward way.

In terms of tackling estimation using this prior we note that adding the prior \eqref{dSSVS1} to our benchmark TVP specification introduces some peculiarity:  by combining equations \eqref{TVP_state} and \eqref{dSSVS1} we end up having two conditional prior structures for $\beta_{j,t}$, namely 
\begin{eqnarray}
\beta_{j,t} \vert \beta_{j,t-1}, w_{j,t} &\sim& N \left(\beta_{j,t-1}, w_{j,t}\right)  \label{beta_prior_1}\\
\beta_{j,t} \vert \gamma_{j,t},\tau_{j,t}^{2} &\sim& N \left( 0,v_{j,t} \right),   \label{beta_prior_2}
\end{eqnarray}
where we define $ v_{j,t} = \left( 1 - \gamma_{j,t} \right)^{2}\underline{c} \times \tau_{j,t}^{2} + \gamma_{j,t}^{2} \tau_{j,t}^{2} $ and $V_{t}$ is the $p \times p$ diagonal matrix comprising the elements $v_{j,t}$. Following ideas in \cite{Wangetal2016} we combine the two priors for $\beta_{t}$ described above by rewriting the state equation as\footnote{\label{foot:wang}The derivation is straightforward using arguments in the previous subsection, see equation \eqref{third_part}. Define $q \left( \beta_{t} \vert y_{1:t-1} \right)$ to be the time $t$ variational Bayes prior of $\beta_{t}$ given information at time $t-1$. Then we have
\begin{eqnarray}
q \left( \bm \beta_{t} \vert \bm y_{1:t-1}  \right) &\propto & \exp \left\lbrace \mathbb{E} \left( \log p\left( \bm \beta_{t} \vert \bm \beta_{t-1}, \bm W_{t} \right) \right) + \mathbb{E} \left( \log p \left( \bm \beta_{t} \vert \bm V_{t} \right) \right) \right\rbrace \nonumber \\
 &\propto & \exp \left\lbrace  - \frac{1}{2} \left(\bm \beta_{t} - \bm \beta_{t-1} \right)^{\prime} \bm W_{t}^{-1}\left(\bm \beta_{t} - \bm \beta_{t-1} \right)  - \frac{1}{2} \bm \beta_{t}^{\prime} \bm V_{t}^{-1} \bm \beta_{t} \right\rbrace  \nonumber \\
  &\propto & \exp \left\lbrace  - \frac{1}{2} \bm \beta_{t}^{\prime} \bm W_{t}^{-1}\bm \beta_{t} + \bm \beta_{t}^{\prime} \bm W_{t}^{-1} \bm \beta_{t-1} - \frac{1}{2} \bm \beta_{t}^{\prime} \bm V_{t}^{-1} \bm \beta_{t} \right\rbrace  \nonumber \\
 &\propto & \exp \left\lbrace - \frac{1}{2} \left( \bm \beta_{t} - \widetilde{\bm F}_{t} \bm \beta_{t-1} \right)^{\prime} \widetilde{\bm W}_{t}^{-1} \left( \bm \beta_{t} - \widetilde{\bm F}_{t} \bm \beta_{t-1} \right)  \right\rbrace , \nonumber 
\end{eqnarray}
where the simplification occurs due to the fact that $\bm \beta_{t-1}$ is known and fixed (i.e. not a random variable) given information at time $t-1$. Therefore, the formula above specifies the new, joint time $t$ prior of $\bm \beta_{t}$ given the two priors in equations \eqref{beta_prior_1}-\eqref{beta_prior_2}.} 
\begin{equation}
\bm \beta_{t} = \widetilde{\bm F}_{t} \bm \beta_{t-1} + \widetilde{\bm \eta}_{t}, \label{new_ss}
\end{equation}
where $\widetilde{\bm \eta}_{t} \sim N \left( \bm 0, \widetilde{\bm W}_{t} \right)$, with parameter matrices $\widetilde{\bm W}_{t} = \left( \mathbb{E} \left(\bm W_{t}\right)^{-1} + \mathbb{E} \left(\bm V_{t}\right)^{-1} \right)^{-1}$ and $\widetilde{\bm F}_{t} = \widetilde{\bm W}_{t} \times \mathbb{E} \left(\bm W_{t}\right)^{-1}$, where $\bm W_{t} = diag \left(w_{1,t},...,w_{p,t} \right)$ and $\bm V_{t} = diag \left(v_{1,t},...,v_{p,t} \right)$, and all expectation operators are with respect to $q\left( \bm \beta_{t} \vert \bm y_{1:t}\right)$. Under this formulation we can observe that the joint prior variance for $\beta_{j,t}$ is a function of both $w_{j,t}$ and $v_{j,t}$, $\forall j=1,...,p$. Therefore, the TVP regression model with dynamic variable selection prior can be written using a new state-space form, with measurement equation given by \eqref{TVP_measurement} and state equation given by \eqref{new_ss}.

Application of algorithm \autoref{algorithm:VBKF_offline} to the transformed state-space model consisting of equations \eqref{TVP_measurement_new} and \eqref{new_ss} provides as output estimates $\bm m_{t\vert T}$ $\forall t$, that is, the smoothed posterior mean of $q \left( \bm \beta_{t} \vert y_{1:T} \right)$. Conditional on these estimates, derivation of the update steps for $\gamma_{j,t}$, $\tau^{2}_{j,t}$ and $\pi_{0,t}$ relies also on deriving the expectations of these variables with respect to $q \left( \bm \beta_{t} \vert y_{1:T} \right)$. Therefore, extending the analysis of the previous section to accommodate these new parameters, and similar to derivations found in Gibbs sampling approaches to variable selection \citep[see, for instance, the formulas of the conditional posteriors in][]{GeorgeMcCulloch1993}, the updating steps for the parameters in the dynamic variable selection prior are the following
\begin{eqnarray}
\widehat{\tau}_{j,t}^{2} = \mathbb{E} \left[ q \left(\tau_{j,t}^{2} \vert y_{t}\right) \right] & = & \left( h_{0} + m_{j,t|T}^{2} \right) /\left(g_{0} + 1/2 \right), \\
\widehat{\gamma}_{j,t} = \mathbb{E} \left[ q \left( \gamma_{j,t} \vert y_{t} \right) \right] & = & \frac{N\left(m_{j,t \vert T} \vert 0,\widehat{\tau}_{j,t}^{2} \right) \widehat{\pi}_{0,t}}{N\left(m_{j,t \vert T} \vert 0,\widehat{\tau}_{j,t}^{2} \right) \widehat{\pi}_{0,t} + N\left( m_{j,t \vert T \vert t} \vert 0,\underline{c}\times \widehat{\tau}_{j,t}^{2} \right) \left( 1- \widehat{\pi}_{0,t} \right)}, \\
\widehat{v}_{j,t} = \mathbb{E} \left[ q \left( v_{j,t} \vert y_{t} \right) \right] &= & \left( 1- \widehat{\gamma}_{j,t} \right)^{2} \underline{c} \widehat{\tau}_{j,t}^{2} + \widehat{\gamma}_{j,t} \widehat{\tau}_{j,t}^{2}, \\
\widehat{\pi}_{0,t} = \mathbb{E} \left[ q \left( \pi_{0,t} \vert y_{t} \right) \right] &=& \left( 1 + \sum_{j=1}^{p} \widehat{\gamma}_{j,t} \right) / (2+p),
\end{eqnarray}
for each $t=1,..,T$ and $j=1,...,p$, where again expectations $\mathbb{E}$ are with respect to the VB posteriors of each of the parameters showing up on the right-hand side of the equations above.

\subsection{Adding stochastic volatility} \label{sec:sv}
A known regression variance is far from a realistic assumption for most datasets. When forecasting macroeconomic data, so is the assumption of an unknown variance that is constant over time. A vast recent literature highlights the importance of time-varying volatility in improving point and density forecasts \citep{ClarkRavazzolo2015}, and the purpose of this subsection is to accommodate estimation of the parameter $var(\varepsilon_{t}) = \sigma_{t}^2$ in the VB setting. Several elegant algorithms for VB inference in stochastic volatility models exist in the literature. For example, \cite{Naessethetal2017} introduce a variational Bayes Sequential Monte Carlo (SMC) algorithm for stochastic volatility models. \cite{Tranetal2017} propose a variational Bayes method for intractable likelihoods that does not rely on the mean field approximation, and apply their algorithm to the estimation of a stochastic volatility model.

Nevertheless, such algorithms assume an explicit time-series model for the stochastic volatility parameter, an assumption that is only useful in a setting where one is interested in forecasting volatility. In a macroeoconomic setting we are interested in forecasting $y_{t}$ and not its volatility (as it would be the case in empirical asset pricing). At the same time, previous empirical work shows that there are no statistically important differences when forecasting with alternative specifications of macroeconomic volatility.\footnote{For example, \cite{ClarkRavazzolo2015} compare a range of specifications for time-varying variance parameters in univariate and multivariate autoregressive models, and any differences among such specifications are not statistically important (while all volatility specifications are always better relative to constant variance specifications).} For that reason, our aim here is not only to render estimation of stochastic volatility precise, but at the same time numerically reliable and computationally efficient. In order to achieve this, we build on variance discounting ideas for dynamic linear methods as described in \cite{WestHarrison1997}; see also \cite{RockovaMcAlinn2017}.

Define $\phi_{t} = \frac{1}{\sigma_{t}^{2}}$ to be the precision (inverse variance). Following \cite{WestHarrison1997} we assume that the time $t-1$ posterior of $\phi$ has the following conjugate form
\begin{equation}
\phi_{t-1} \vert y_{1:t-1} \sim Gamma \left( a_{t-1},b_{t-1} \right).
\end{equation}
We do not specify an explicit time series model for the dynamics of $\phi$ (e.g. stochastic volatility or GARCH) because the posterior for $\phi_{t}$ wouldn't be conjugate to the likelihood and we would fail to obtain fast updates. In order to maintain this conjugacy we specify instead the time $t$ prior of the form
\begin{equation}
\phi_{t} \vert y_{1:t-1} \sim Gamma \left( \delta a_{t-1},\delta b_{t-1} \right), \label{phi_prior}
\end{equation}
for a variance discounting factor $0<\delta<1$, subject to a choice of hyperparameters $a_{0}$ and $b_{0}$. By doing so, we assume that $\phi_{t}$ is centered around $\phi_{t-1}$ as if this parameter had random walk dynamics,\footnote{Even though we haven't specified an explicit time series evolution for $\phi_{t}$, by using results in \cite{Uhlig1994} we can show that the proposed variance discounting methodology is equivalent to assuming the following specification:
\begin{equation}
\phi_{t} = \gamma_{t} \phi_{t-1}/ \delta,
\end{equation}
for a parameter $\gamma_{t} \vert y_{1:t-1} \sim Beta \left( \delta a_{t-1}/2, (1-\delta) a_{t-1} /2 \right)$.} since it holds that $ \mathbb{E} \left( \phi_{t} \vert y_{1:t-1} \right)= \mathbb{E} \left( \phi_{t-1} \vert y_{1:t-1} \right)$. However, based on the properties of the Gamma distribution, the dispersion of $\phi_{t}$ is larger to that of $\phi_{t-1}$.

Under this scheme the variational Bayes update of $\phi_{t}$, that is, its time $t$ posterior mean has the form
\begin{equation}
\widehat{\phi}_{t} = \mathbb{E}_{q(\beta_{t} \vert y_{1:T})}\left(\phi_{t} \vert y_{1:t} \right) = a_{t}/b_{t},
\end{equation}
where $a_{t} = 1/2 + \delta a_{t-1}$ and $d_{t} = \frac{1}{2} \left[ \left(y_{t} - \bm x_{t} \bm m_{t \vert T}) \right)^{2} + \bm x_{t} \bm P_{t \vert T} \bm x_{t}^{\prime} \right] + \delta b_{t-1}$, where $\bm m_{t \vert T},\bm P_{t \vert T}$ are the smoothed mean and variance of $\beta_{t}$. Using this scheme, past information in the data is discounted exponentially by the factor $\delta$. The scalar $\delta$ can be seen as a prior hyperparameter whose choice determines how much relative weight we give to recent versus older observations, that is, it determines how fast we expect the precision parameter to change over time. For $\delta=1$ we obtain the posterior under a standard recursive update scheme (similar to recursive OLS), while typical values that would allow for faster time-variation in the precision/variance would be between 0.8 and 0.99. Values lower than 0.8 are not empirically advised, since they allow for a large amount of time-variation and stochastic variance estimates become very noisy. In the empirical exercise we set $\delta=0.8$, a choice that reflects our prior expectation that macroeconomic data have many abrupt breaks in their second moments and excess kurtosis during recessions (implying variances that can move very fast over time).

The previous formulas pertain to the iterative updating of $\phi_{t}$ given $\phi_{t-1}$. Estimates of $\phi_{t}$ can be smoothed using subsequent observations $t+1,...,T$. Following \cite{WestHarrison1997} we can run a backward recursive filter of the form
\begin{equation}
\widetilde{\phi}_{t} = (1-\delta)\widehat{\phi}_{t} +\delta\widetilde{\phi}_{t+1}, 
\end{equation}
for $t=T-1,...,1$, where $\widetilde{\phi}_{t} = \mathbb{E}_{q(\beta_{t} \vert y_{1:T})}\left(\phi_{t} \vert y_{t+1} \right)$ and $\widetilde{\phi}_{T} = \widehat{\phi}_{T}$. Once we obtain this update for the precision $\phi_{t}$, a posterior mean estimate of the volatility $\sigma_{t}^{2}$ can be obtained simply as the inverse of $\widetilde{\phi}_{t}$.
\subsection{The Variational Bayes Dynamic Variable Selection (VBDVS) algorithm}
Here we provide details of the exact parameter updates that result from VB inference in our proposed specification. Algorithm \autoref{algorithm:VBDVS} outlines our proposed \emph{Variational Bayes Dynamic Variable Selection (henceforth, VBDVS)} algorithm. This Algorithm shows an accurate picture of how this would look like when programmed using a language like MATLAB or R: while there are many parameters involved in our specification, the code is short and it involves simple scalar operations (meaning it is very fast). The only cumbersome operation is the inversion of the $p \times p$ matrix $\bm P_{t+1|t}$ in line 14 which has worst case complexity $\mathcal{O}\left(p^3\right)$ for each $t$. There are four main blocks in this algorithm. Lines 4-12 are a result of straightforward application of the Kalman filter on the state-space model of equations \eqref{TVP_measurement_new} and \eqref{new_ss}, and lines 13-17 show the backwards (smoothing) recursions. Lines 18-27 update the prior hyperparameters of the DVS prior for $\bm \beta_{t}$. Finally, lines 28-33 provide updates for the stochastic volatility parameter, as discussed in the previous subsection.

\begin{algorithm}[htbp!]
\caption{\textit{Variational Bayes algorithm for TVP regression model with dynamic variable selection and stochastic variance (VBDVS algorithm)}}\label{algorithm:VBDVS}
  \scriptsize
\begin{algorithmic}[1]
\State Choose values of $\bm m_{0},\bm P_{0},a_0,b_0,c_{j,0},d_{j,0},g_0,h_0,\underline{c}$, and $\delta$; initialize all vectors/matrices.
\State $\mathcal{r}=1$
\While{$ \Vert \mathcal{G}\left(q(\bm \beta^{(\mathcal{r})},\bm w^{(\mathcal{r})} \vert \bm y \right) - \mathcal{G}\left(q(\bm \beta^{(\mathcal{r}-1)},\bm w^{(\mathcal{r}-1)} \vert \bm y \right) \Vert \rightarrow 0$}
    \For{$t = 1$ \algorithmicto $T$}
         \State $\widetilde{\bm W}_{t}^{\left(\mathcal{r}\right)} = diag\left( \left( w_{1,t}^{-1 \text{ \ } \left(\mathcal{r}-1\right)} + v_{1,t}^{-1 \text{ \ } \left(\mathcal{r}-1\right)}\right)^{-1},...,\left(w_{p,t}^{-1 \text{ \ } \left(\mathcal{r}-1\right)} +  v_{p,t}^{-1 \text{ \ } \left(\mathcal{r}-1\right)} \right)^{-1}\right)$
          \State $\widetilde{\bm F}_{t}^{\left(\mathcal{r}\right)} = \widetilde{\bm W}_{t}^{\left(\mathcal{r}\right)}\left(\bm W_{t}^{\left(\mathcal{r}-1\right)}\right)^{-1}$
         \State $\bm m_{t|t-1}^{\left(\mathcal{r}\right)} = \widetilde{\bm F}_{t}^{\left(\mathcal{r}\right)} \bm m_{t-1 \vert t-1}^{\left(\mathcal{r}\right)}$   \hfill  \texttt{Predicted mean}
         \State $\bm P_{t|t-1}^{\left(\mathcal{r}\right)} = \widetilde{\bm F}_{t}^{\left(\mathcal{r}\right)} \bm P_{t-1 \vert t-1}\widetilde{\bm F}_{t}^{\left(\mathcal{r}\right)^{\prime}} + \widetilde{\bm W}_{t}^{\left(\mathcal{r}\right)}$  \hfill  \texttt{Predicted variance}
         \State $\bm K_{t}^{\left(\mathcal{r}\right)} = \bm P_{t|t-1}^{\left(\mathcal{r}\right)} \bm x_{t}^{\prime} \left( \bm x_{t} \bm P_{t|t-1}^{\left(\mathcal{r}\right)} \bm x_{t}^{\prime} + \widehat{\sigma}^{2 \text{ \ } \left(\mathcal{r}-1\right)}_{t} \right)^{-1} $   \hfill  \texttt{Kalman gain}
         \State $\bm m_{t|t}^{\left(\mathcal{r}\right)} = \bm m_{t|t-1}^{\left(\mathcal{r}\right)} + \bm K_{t}^{\left(\mathcal{r}\right)} \left(y_{t} - \bm x_{t} \bm m_{t|t-1}^{\left(\mathcal{r}\right)} \right)$    \hfill  \texttt{Filtered mean of $\bm \beta_{t}$}
         \State $\bm P_{t|t}^{\left(\mathcal{r}\right)}  =  \left( \bm I_{p} - \bm K_{t}^{\left(\mathcal{r}\right)} \bm x_{t} \right) \bm P_{t|t-1}^{\left(\mathcal{r}\right)}$ \hfill  \texttt{Filtered variance of $\bm \beta_{t}$}
         \EndFor
         \For{$T = T-1$ \algorithmicto $1$}
         \State $\bm C  = \bm P_{t|t}^{\left(\mathcal{r}\right)} \widetilde{\bm F}_{t}^{\left(\mathcal{r}\right)} \left( \bm P_{t+1|t}^{\left(\mathcal{r}\right)} \right)^{-1}$
         \State $\bm m_{t|T}^{\left(\mathcal{r}\right)} = \bm m_{t|t}^{\left(\mathcal{r}\right)} + \bm C \left(\bm m_{t+1|T}^{\left(\mathcal{r}\right)} - \bm m_{t+1|t}^{\left(\mathcal{r}\right)} \right)$    \hfill  \texttt{Smoothed mean of $\bm \beta_{t}$}
         \State $\bm P_{t|T}^{\left(\mathcal{r}\right)} = \bm P_{t|t}^{\left(\mathcal{r}\right)} + \bm C \left( \bm P_{t+1|T}^{\left(\mathcal{r}\right)} - \bm P_{t+1|t}^{\left(\mathcal{r}\right)}\right) \bm C^{\prime}$   \hfill  \texttt{Smoothed variance of $\bm \beta_{t}$}
         \EndFor
        \State $\bm D_{t} = \bm P_{t \vert T}^{\left(\mathcal{r}\right)} + \bm m_{t|T}^{\left(\mathcal{r}\right)} \bm m_{t|T}^{\left(\mathcal{r}\right)^{\prime}} + \left( \bm P_{t-1|T}^{\left(\mathcal{r}\right)} + \bm m_{t-1|T}^{\left(\mathcal{r}\right)} \bm m_{t-1|T}^{\left(\mathcal{r}\right)^{\prime}} \right) \left( I_{p} - 2\widetilde{F}_{t}^{\left(\mathcal{r}\right)} \right)^{\prime}$ \hfill \texttt{Squared error in state eq.}
        \State $R_{t} = \left[ \left( y_{t} - \bm x_{t} \bm m_{t|T}^{\left(\mathcal{r}\right)}  \right)^2 + \bm x_{t} \bm P_{t|T} \bm x_{t}^{\prime} \right] $ \hfill \texttt{Squared error in measurement eq.}
        \For{$t = 1$ \algorithmicto $T$}
        \For{$j = 1$ \algorithmicto $p$}
            \State $\widehat{\tau}_{j,t}^{-2 \text{ \ } \left(\mathcal{r}\right)} =  \left(g_{0} + 0.5 \right) / \left( h_{0} + 0.5\left(m_{j,t|T}^{\left(\mathcal{r}\right)}\right)^{2} \right)  $ \hfill \texttt{Posterior mean of $\frac{1}{\tau_{j,t}^{2}}$}
            \State $\widehat{\gamma}_{j,t}^{\left(\mathcal{r}\right)} = \frac{N\left(m_{j,t \vert T}^{\left(\mathcal{r}\right)} \vert 0,\widehat{\tau}_{j,t}^{2 \text{ \ } \left(\mathcal{r}\right)} \right) \widehat{\pi}_{0,t}^{\left(\mathcal{r}-1\right)}}{N\left(m_{j,t \vert T}^{\left(\mathcal{r}\right)} \vert 0,\widehat{\tau}_{j,t}^{2 \text{ \ } \left(\mathcal{r}\right)} \right) \widehat{\pi}_{0,t}^{\left(\mathcal{r}-1\right)} + N\left( m_{j,t \vert T \vert t}^{\left(\mathcal{r}\right)} \vert 0,\underline{c}\times \widehat{\tau}_{j,t}^{2 \text{ \ } \left(\mathcal{r}\right)} \right) \left( 1- \widehat{\pi}_{0,t}^{\left(\mathcal{r}-1\right)} \right)}$\hfill \texttt{Posterior mean of $\gamma_{j,t}$}  
           \State $ \widehat{v}_{j,t}^{\left(\mathcal{r}\right)} = \left( 1- \widehat{\gamma}_{j,t}^{\left(\mathcal{r}\right)} \right)^{2} \underline{c} \widehat{\tau}_{j,t}^{2 \text{ \ } \left(\mathcal{r}\right)} + \widehat{\gamma}_{j,t}^{\left(\mathcal{r}\right)} \widehat{\tau}_{j,t}^{2 \text{ \ } \left(\mathcal{r}\right)} $\hfill \texttt{Posterior mean of $v_{j,t}$}            
            \State $\widehat{w}_{j,t}^{-1 \text { \ } \text{ \ }\left(\mathcal{r}\right)} =  \left(\underline{c}_{0} + 0.5 \right) / \left( \underline{d}_{0} + 0.5 \bm D_{jj,t} \right)  $ \hfill \texttt{Posterior mean of $\frac{1}{w_{j,t}}$}
        \EndFor
        \State $\widehat{\pi}_{0,t}^{\left(\mathcal{r}\right)} = \left( 1 + \sum_{j=1}^{p} \widehat{\gamma}_{j,t}^{\left(\mathcal{r}\right)} \right) / (2+p)$ \hfill \texttt{Posterior mean of $\pi_{0,t}$}
        \State $\widehat{\phi}_{t}^{\left(\mathcal{r}\right)} = (\delta a_{t-1} + 0.5)/ (\delta b_{t-1} + 0.5R_{t})$ \hfill \texttt{Filtered mean of $\frac{1}{\sigma_{t}^{2}}$}
        \EndFor
        \For{$T = T-1$ \algorithmicto $1$}
        \State $\widetilde{\phi}_{t}^{\left(\mathcal{r}\right)} = (1-\delta) \widehat{\phi}_{t}^{\left(\mathcal{r}\right)} + \delta \widetilde{\phi}_{t+1}^{\left(\mathcal{r}\right)}$ \hfill \texttt{Smoothed mean of $\frac{1}{\sigma_{t}^{2}}$}
        \EndFor        
        \State $\mathcal{r} = \mathcal{r} + 1$
    \EndWhile
\end{algorithmic}
\end{algorithm}

\section{Simulation study}
In this section we evaluate the performance of the new estimator using artificial data. Although we view the algorithm as primarily a forecasting algorithm, it is also important to investigate its estimation accuracy in an environment where we know the true data generating process (DGP). Thus, we wish to to establish that the VBDVS is able to track time-varying parameters satisfactorily and establish that the dynamic variable selection prior is able to perform shrinkage and selection with high accuracy (at least in cases where we know that the DGP is that of a sparse TVP regression model). We also wish to investigate the computational gains that arise from application of variational Bayes methods on the complex dynamic variable selection prior structure.

In all our experiments we use the following DGP: 
\begin{eqnarray}
y_{t} & = & \beta_{1t}x_{1t} + \beta_{2t}x_{2t} + ... + \beta_{pt}x_{pt} + \sigma_{t}\varepsilon_{t}, \text{   } \varepsilon_{t} \sim N(0,1) \\
x_{j,t} & \sim & N(0,1), \text{  } j=1,...,p \\
\beta_{j,t} & = & s_{j,t} \times \theta_{j,t} \\
\theta_{j,t} & = & \underline{\theta}_{j} +  \underline{\rho}\left( \theta_{j,t-1} - \underline{\theta}_{j} \right)+ \underline{\delta} \eta_{j,t},  \text{   } \eta_{j,t} \sim N(0,1) \label{TVP_DGP} \\
\log \left( \sigma_{t}^{2} \right) & = & \underline{\sigma}^{2} +   \underline{\phi} \left( \log \left( \sigma_{t-1}^{2} \right) - \underline{\sigma}^{2} \right) + \underline{\xi} \zeta_{t}, \text{   } \zeta_{t} \sim N(0,1) \label{SV_DGP} \\
\theta_{j,0} & = & \underline{\theta}_{j}, \text{ \ \ \ \ } \log \left(\sigma_{0}^{2}\right) = \underline{\sigma}^{2}.
\end{eqnarray}
Our benchmark specification sets $\underline{\bm \theta} = \left(-1.7, 2.9, 1.4, -2.3, \bm 0 \right)$, $\underline{\sigma}^{2} = 0.1$, $\underline{\rho} = \underline{\phi} = 0.99$, $\underline{\delta} = \underline{\xi}= T^{-1/2}$. In the specification above $\bm s_{j}$ is $T \times 1$ vector of either zeros or ones, such that $\beta_{j,t}=\theta_{j,t}$ when $s_{j,t}=1$, and zero otherwise.  We set $s_{1,t}=1$ for $t=1,...,\lfloor T/3 \rceil-1$ and zero otherwise, $s_{2,t}=1$ $\forall t=1,...,T$, $s_{3,t}=1$ for $t=1,...,\lfloor T/2 \rceil-1$ and zero otherwise, $s_{4,t} = 0$ for $t=1,...,\lfloor T/2 \rceil-1$ and zero otherwise. These choices mean that $\beta_{1,t}$ is zero during the last third of the sample, $\beta_{2,t}$ is a relevant predictor in all periods, $\beta_{3,t}$ is zero during the last half of the sample, and $\beta_{4,t}$ is zero during the first half of the sample. Any other coefficient for $j=5,...,p$ is zero at all periods, i.e. $s_{j,t}=0$ $\forall$ $j>4$, $t=1,...,T$. By doing so, we simulate a situation where only one predictor is relevant in all time periods, three predictors are relevant only in certain subsamples of the data, and all remaining $p-4$ predictors are irrelevant for $y$ at all time periods.

After we generate artificial data, we compare three competing estimation algorithms for TVP models: i) our variational Bayes dynamic variable selection (VBDVS) algorithm, ii) the EM algorithm implementation of the dynamic spike and slab (DSS) of \cite{RockovaMcAlinn2017}, and iii) Gibbs sampling (MCMC) estimation of the TVP model using the fast algorithm of \cite{ChanJeliazkov2009}. While there are numerous other algorithms available for estimating TVP models, our limited choice of algorithms reflects our desire to simulate exclusively high-dimensional models. By doing so, we exclude most of the recently proposed Bayesian methodologies cited in the Introduction. These methodologies introduce various flexible parametrizations (like we do) that result, however, in the need for many tuning parameters and estimation via MCMC, such that they become unreasonably cumbersome for $p>50$. Our model instead, as we demonstrate in detail later, requires very straightforward tuning. The default prior setting we use for the VBDVS algorithm is based on the case Prior 3 presented in \autoref{table:priors} in the next section. The settings used in the DSS and MCM algorithms are discussed in the Online Supplement to this paper. In order to compare numerically these algorithms we generate $R=100$ datasets from the above DGP for various choices of sample size and total number of predictors, namely $T=100,200,500$ and $p=50,100,200$. Subsequently squared deviations between true and estimated parameters are calculated, and then averaged over the $T$ time periods, and $p$ predictors. To be precise, if we let $\left( \beta_{t}^{true} \right)$ denote the true artificially generated coefficients and $\left( \beta_{t}^{j}, \sigma_{t}^{j} \right)$, for $j=DVS,DSS,MCMC$, the estimates of these coefficients, we calculate the sum of mean squared deviations (MSD) statistic as
\begin{eqnarray}
MSD_{\beta}^{j} & = & \sum_{r=1}^{100}  \left( \frac{1}{T \times p} \sum_{t=1}^{T} \sum_{i=1}^{p} \left( \beta_{it}^{true, (r)} - \beta_{it}^{j, (r)}   \right)^{2} \right),
\end{eqnarray}
where $r=1,...,500$ denotes the number of Monte Carlo iterations.

\begin{figure}[H]  
\centering
\includegraphics[scale=0.5,trim={3cm 1cm 3cm 1cm}]{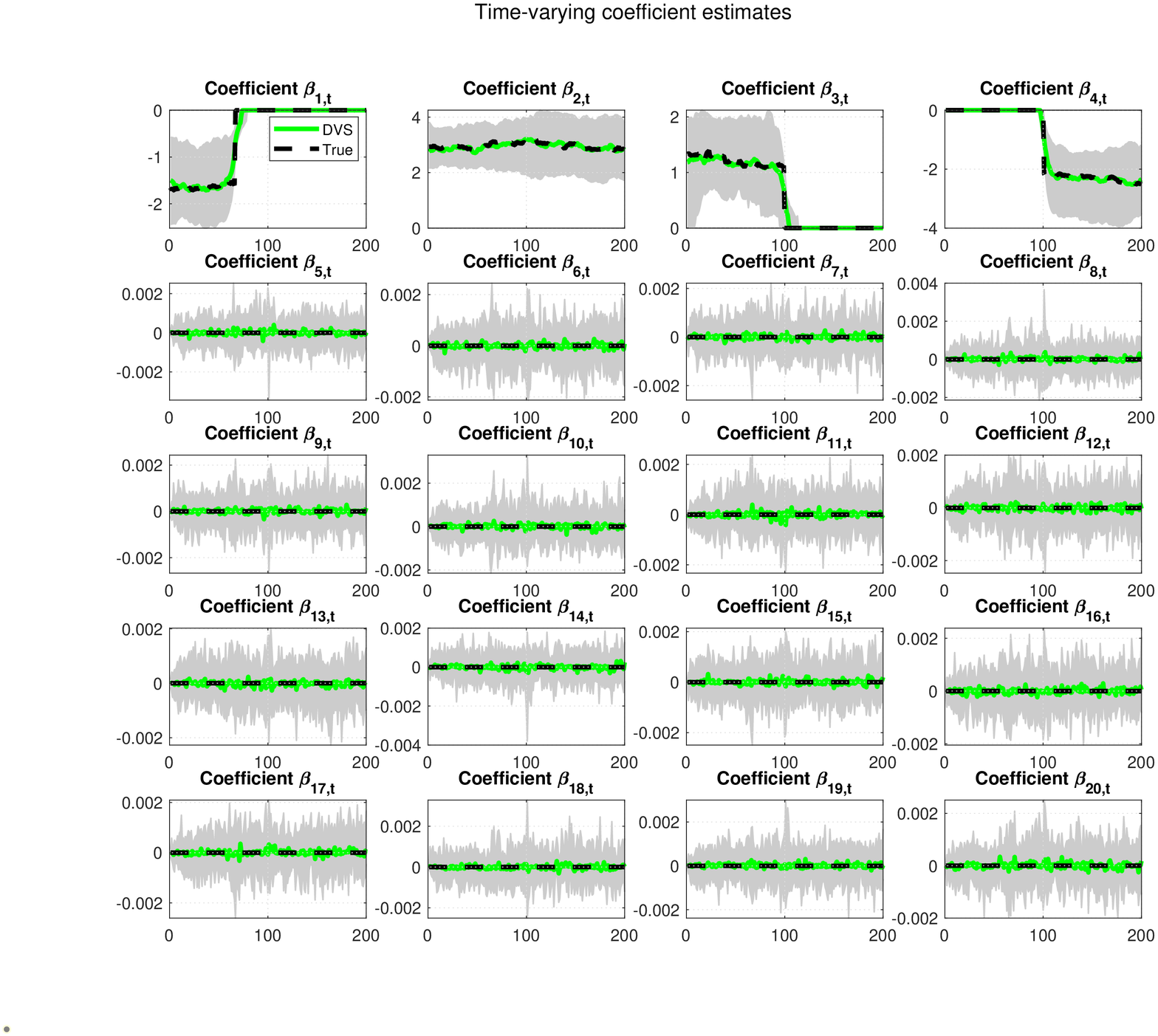}
\caption{\textit{VBDVS coefficient estimates of the first 20 predictors generated from the DGP with $T=200$ and $p=200$. Black dashed lines are the true generated coefficients. Posterior medians (over the 100 Monte Carlo iterations) of VBDVS estimates are shown with green solid lines, and grey areas denote 16$^{th}$ and 84$^{th}$ percentiles.}}  \label{fig:MC1}
\end{figure}
 
\begin{figure}[H]  
\centering
\includegraphics[scale=0.5,trim={3cm 1cm 3cm 1cm}]{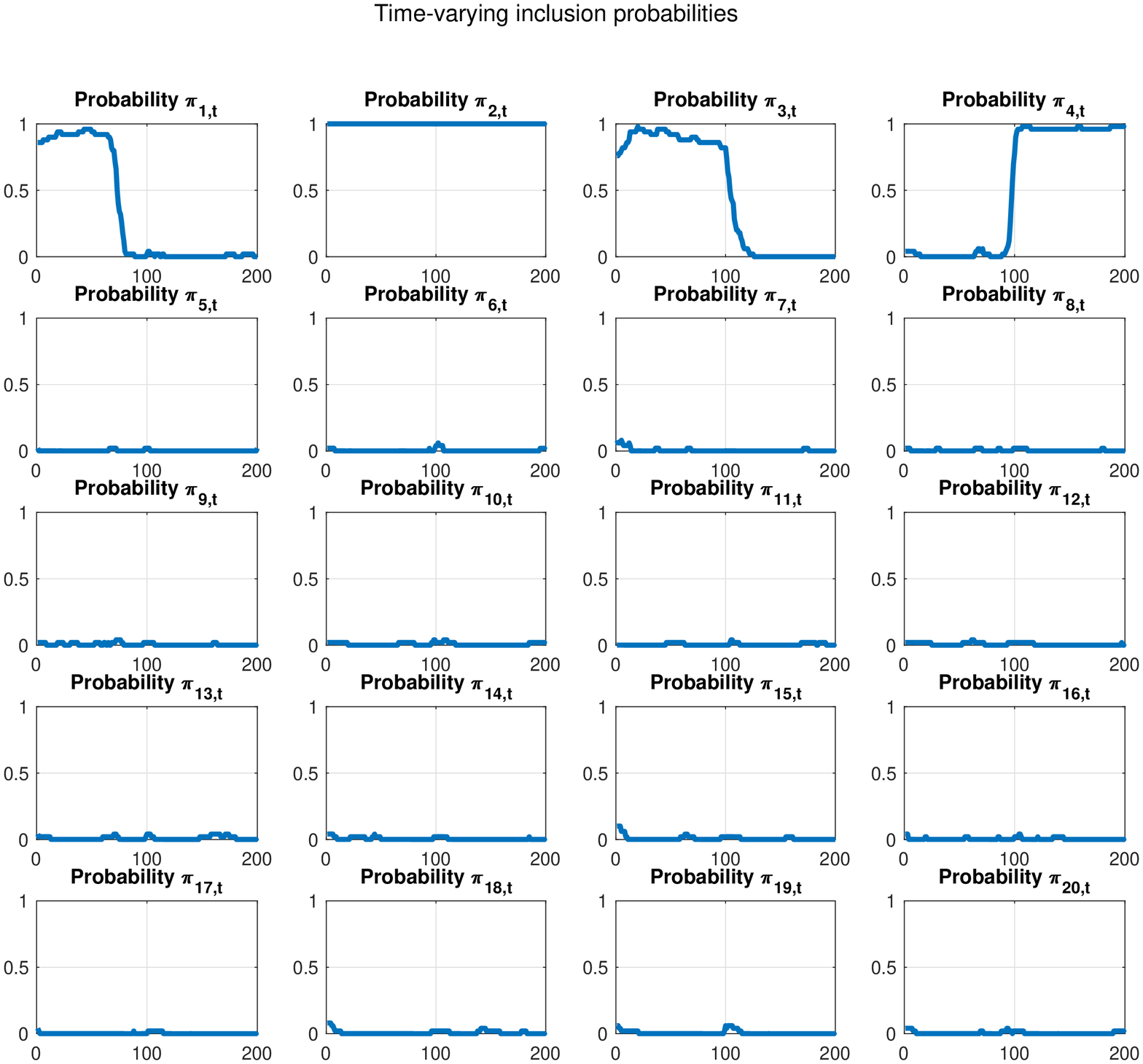}
\caption{\textit{Time-varying posterior inclusion probabilities (expected value of $\gamma_{j,t}$ estimates) of the first 20 predictors generated from the DGP with $T=200$ and $p=200$. These probabilities are means over the 100 Monte Carlo iterations.}}  \label{fig:MC2}
\end{figure}

\autoref{fig:MC1} shows the coefficient estimates from VBDVS for the case $T=p=200$. This plot compares the posterior median (green solid lines) versus the true generated coefficients (black dashed lines). The 16$^{th}$ and 84$^{th}$ percentiles over the 100 Monte Carlo iterations are also shown as a shaded area around the posterior median. Only the first 20 coefficients, out of the possible 200, are plotted. The first row shows the four coefficients that, at least in some periods, are non-zero, followed by 16 coefficients that are exactly zero. It is impossible to plot the remaining 180 coefficients in the DGP that are exactly zero, but their estimates are represented fairly well by the estimates of coefficients $\beta_{5,t}$ - $\beta_{20,t}$ shown in \autoref{fig:MC1}. Under the assumption of sparsity in the DGP, the VBDVS algorithm is able to recover the true coefficients with accuracy. Not only the coefficients that are zero in the DGP in all periods are correctly estimated to be zero, but also the three coefficients that are zero only in certain subsamples are estimated precisely. When a coefficient is initially zero and later in the sample becomes important (see coefficient $\beta_{4,t}$), and vice-versa (see coefficients $\beta_{1,t}$ and $\beta_{3,t}$), the dynamic variable selection algorithm is able to identify and jump quickly to the new state. \autoref{fig:MC2} shows that the true reason why estimation is so precise -- even in such a demanding case with 200 time-varying coefficients for only 200 observations -- is because the estimates of the time-varying posterior inclusion probabilities (PIPs) of each predictor are recovered with precision in the first instance. By identifying correctly which variables should be excluded from the regression model in each period results in shrinking many coefficients to zero and allowing to preserve enough degrees of freedom for estimation of non-zero coefficients.

\autoref{table:MC_results} shows the values of the MSD statistics for the three algorithms under the different combinations of $T$ and $p$. Given that the MSD statistics measure deviation from the true coefficient, lower values imply that a certain estimation algorithm has done better recovery of the coefficients generated by the DGP. In all cases VBDVS has the best performance among all competing algorithms. The estimation error of the MCMC algorithm is quite large mainly because the algorithm is unable to shrink all $p-4$ coefficients in the DGP that are exactly zero. The DSS algorithm provides a better fit since it is also an algorithm that does dynamic variable selection and shrinkage. Its performance is slightly inferior to VBDVS, but the results should not be taken as final evidence. While we have done all effort to follow the settings suggested by \cite{RockovaMcAlinn2017}, there might be other priors that could improve the performance of this algorithm. 

Another important feature of the VBDVS algorithm is its fast computing time. While it is not surprising that our algorithm is faster compared to MCMC, our algorithm can provide substantial savings in high-dimensional settings compared to the DSS that relies on the EM algorithm. Columns 6-8 in \autoref{table:MC_results} reveals that VBDVS can be multiple times faster than both DSS and MCMC algorithms.
  
\begin{table}[H]
\begin{center}
\caption{\textit{MSD statistics and computing time for Monte Carlo exercise}}  \label{table:MC_results}
\resizebox{0.85\textwidth}{!}{
\begin{tabular}{llrrrrrrr} \hline
&	&	\multicolumn{3}{c}{MSD statistic}					&		&	\multicolumn{3}{c}{Computing time (secs)}					\\ \hline
&	&	VBDVS	&	DSS	&	MCMC	&		&	VBDVS	&	DSS	&	MCMC	\\ \hline\hline
	&	$p=50$	&	0.203	&	0.419	&	7.979	&		&	1.2	&	8.3	&	22.6	\\
$T=100$	&	$p=100$	&	0.469	&	1.014	&	11.787	&		&	7.2	&	20.1	&	106.6	\\
	&	$p=200$	&	0.536	&	1.915	&	14.628	&		&	29.9	&	45.8	&	402.0	\\
	&		&		&		&		&		&		&		&		\\
	&	$p=50$	&	0.047	&	0.256	&	5.825	&		&	5.5	&	19.9	&	49.9	\\
$T=200$	&	$p=100$	&	0.088	&	0.789	&	10.583	&		&	10.1	&	40.1	&	232.2	\\
	&	$p=200$	&	0.165	&	1.780	&	17.983	&		&	38.6	&	91.9	&	841.4	\\
	&		&		&		&		&		&		&		&		\\
	&	$p=50$	&	0.019	&	0.147	&	4.613	&		&	8.3	&	51.1	&	125.2	\\
$T=500$	&	$p=100$	&	0.043	&	0.819	&	9.095	&		&	50.9	&	125.1	&	555.6	\\
	&	$p=200$	&	0.085	&	1.679	&	18.398	&		&	83.6	&	220.6	&	2127.8	\\
  \hline	
\end{tabular}
}
\end{center}
{\emph{\scriptsize Notes: Computing times are based on a Windows 10 laptop running MATLAB 2020a, featuring an Intel i7-8665U processor and 32GB of RAM.}}
\end{table}

\section{Macroeconomic Forecasting with Many Predictors}
\subsection{A new large dataset for forecasting inflation} \label{sec:data}
Following a large literature on time-varying parameter models in macroeconomics, our primary target is to forecast quarterly US inflation. While there exists mixed empirical evidence about the potential of very large datasets to improve forecasts of inflation, our aim is to demonstrate here that the new dynamic variable selection methodology can successfully extract, period-by-period, predictive information from a large number of predictors. For that reason we build a novel, high-dimensional dataset that brings together predictors from several mainstream aggregate macroeconomic and financial datasets.\footnote{While one could also think of potential predictors in disaggregated panels obtained in surveys, internet, or documents (text data), such novel sources are typically proprietary and would make our results hard to replicate.} Our building block is the FRED-QD dataset of \cite{McCrackenNg2020}, which we augment with portfolio data used in \cite{Juradoetal2015}, stock market predictors from \cite{WelchGoyal2008}, survey data from University of Michigan consumer surveys, commodity prices from the World Bank's Pink Sheet database, and key macroeconomic indicators from the Federal Reserve Economic Data for four economies (Canada, Germany, Japan, UK). All data are quarterly, and span the period 1960Q1-2018Q4. All variables are adjusted from their respective sources for seasonality (where relevant), and we additionally remove extreme outliers.\footnote{Following \cite{StockWatson2016}, we replace outliers using the median of the preceding five observations. An outlier is defined to be any observation that satisfies $\vert y_{t} - m \vert / iqr >\kappa$, where $m$ is the median of $y$, $iqr $ is the interquantile range, and $\kappa = 4.5$.}

The dataset has in total 444. Out of these we forecast the series (FRED-QD mnemonics in parentheses): GDP deflator (GDPCTPI), total CPI (CPIAUCSL), core CPI (CPILFESL), and PCE deflator (PCECTPI). When each of these price series, $P_{t}$, is used as the dependent variable to be forecasted $h$-quarters ahead we transform it according to the formula $y_{t+h} = (400/h) \ln \left( P_{t+h} - P_{t} \right)$. We forecast these transformed series one at a time, and the remaining three price series are included in the list of exogenous predictor variables (443 in total). The predictor variables are transformed using standard norms in the literature \citep[see for example][]{McCrackenNg2020}: i) levels for variables that are already expressed in rates (e.g. unemployment, interest); ii) first differences of logarithm for variables measuring population (e.g. employment), variables expressed in dollars (e.g. GDP), commodity prices, and some indexes (e.g. Industrial production); and iii) second differences of logarithm for price and consumption indexes, as well as deflator series. The online supplement describes in detail all variables and transformations, and provides links to all sources.

\subsection{How the dynamic variable selection algorithm works: An in-sample assessment}
Before we set up a comprehensive out-of-sample forecasting exercise, we first assess in-sample estimates from the VBDVS by doing small sensitivity analysis to various prior choices. This exercise is intended to demonstrate that the new algorithm provides reasonable estimates of trends, volatilities and other parameters. Most importantly it serves as a way to clarify that, despite the fact that our prior is heavily parametrized, prior elicitation in the VBDVS algorithm becomes a reasonably straightforward task. As it is impossible to present estimates of the TVP model using all variables in our dataset as predictors, we focus on a small TVP model where GDP deflator regressed on an intercept, two own lags, and the first five principal components from the 443 exogenous predictors (eight predictors in total).

Out of all parameters and hyperparameters defined in our algorithm it is only a handful that are crucial for inference and forecasting, while others can be fixed to reasonable or uninformative values and possibly have little effect on forecasting. \autoref{table:priors} lists all hyperparameters one need to choose in the VBDVS algorithm, and does an explicit separation into ``Important'' and ``Fixed'' hyperparameters. Starting from the latter, $a_{0}$ and $b_{0}$ are the initial scale and rate parameters of the initial condition of the precision parameter in equation \eqref{phi_prior}. Setting $a_{0}=b_{0}=0.01$ implies that the precision has prior mean one and variance 10, which is a reasonable uninformative choice for an inverse variance parameter. Next, we set $\delta=0.8$ for reasons explained in \autoref{sec:sv}. Given that $p$ is very large to allow us to obtain meaningful prior information about the regression coefficients $\bm \beta_{t}$ (e.g. using a training sample), we allow their initial condition $\bm \beta_{0}$ to be fairly uninformative by setting $\bm m_{0}= \bm 0$ and $\bm P_{0}=4 \bm I_{p}$. The parameter $\underline{c}$ in the dynamic variable selection prior has to be small (see discussion in \autoref{sec:dvs}) and how small it exactly is, affects the way the algorithm selects each of the two Normal components in the spike and slab prior -- that is, it affects the choice between a certain $\beta_{j,t}$ being restricted or not. We prefer to fix this parameter to $\underline{c}=0.0001$ and allow only $\tau_{j,t}^{2}$ and its prior to determine the ratio of the prior variances of the two Normal components in the mixture prior.

\begin{table}[H]
\begin{center}
\caption{\textit{Hyperparameter choices for sensitivity analysis}} \label{table:priors}
\resizebox{0.5\textwidth}{!}{
\begin{tabular}{lcccl} \hline
	&	Prior 1	&	Prior 2	&	Prior 3	&	Notes	\\ \hline \hline
\multicolumn{4}{l}{\textsc{\underline{Important hyperparameters}}}		\\
$g_{0}$	&	$0.01$	&	$0.01$	&	$1$	&	see eq. \eqref{dSSVS3}	\\
$h_{0}$	&	$0.01$	&	$0.01$	&	$12$	&	see eq. \eqref{dSSVS3}	\\
$c_{j,0}$	&	$100$	&	$1$	&	$100$	&	see eq. \eqref{ssm_prior}	\\
$d_{j,0}$	&	$1$	&	$1$	&	$1$	&	see eq. \eqref{ssm_prior}	\\
\multicolumn{4}{l}{\textsc{\underline{Fixed hyperparameters}}}	\\
$\underline{c}$	&	$10^{-4}$	&	$10^{-4}$	&	$10^{-4}$	&	see eq. \eqref{dSSVS1}	\\
$a_{0}$	&	$0.01$	&	$0.01$	&	$0.01$	&	see eq. \eqref{phi_prior}	\\
$b_{0}$	&	$0.01$	&	$0.01$	&	$0.01$	&	see eq. \eqref{phi_prior}	\\
$\delta$	&	$0.8$	&	$0.8$	&	$0.8$	&	see eq. \eqref{phi_prior}	\\
$ m_{j,0}$	&	$0$	&	$0$	&	$0$	&	see eq. \eqref{ssm_prior}	\\
$ P_{j,0}$	&	$4$	&	$4$	&	$4$	&	see eq. \eqref{ssm_prior}	\\   \hline	
\end{tabular}
}
\end{center}
\end{table}

\begin{figure}[H]  
\centering
\includegraphics[scale=0.35,trim={3cm 1cm 3cm 1cm}]{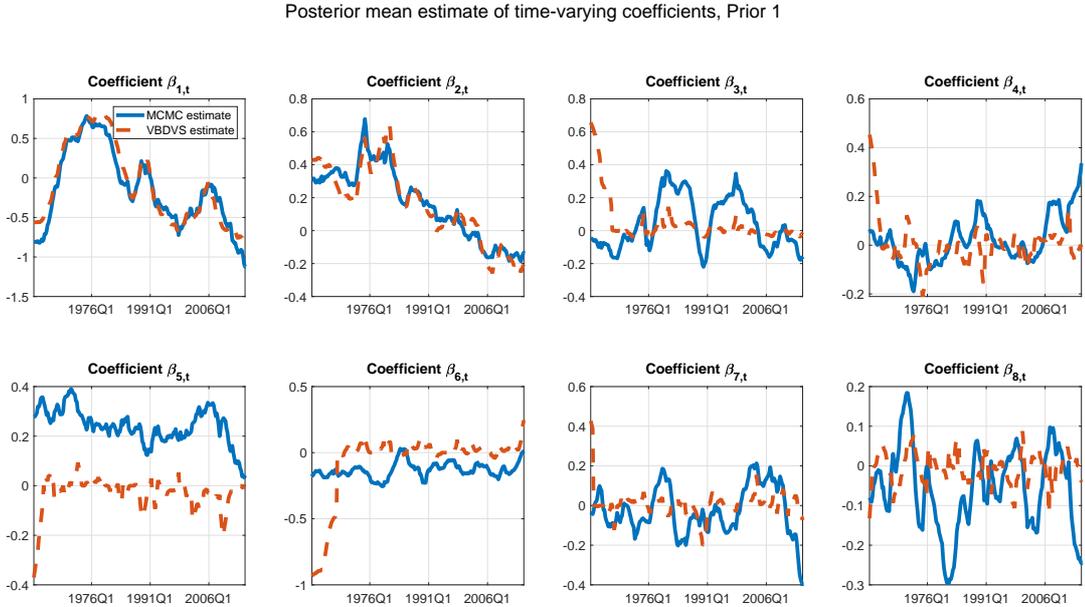}
\caption{\textit{Posterior means of time-varying coefficient estimates from VBDVS (red dashed lines) using Prior 1. Solid lines are posterior means from a TVP model with the same predictors estimated with MCMC.}}  \label{fig:EMP1}
\end{figure}

The parameters that are important in our high-dimensional setting are the ones affecting the two prior variances of the time-varying coefficients $\bm \beta_{t}$, namely the hyperparameters of $\tau_{j,t}^{2}$ and $w_{j,t}$. Our first prior choice, denoted as ``Prior 1'' in \autoref{table:priors}, selects $c_{j,t}=100, d_{j,t}=1$ such that $w_{j,t}$ has a prior mean of 0.01 and prior variance 0.0001. This conservative choice restricts movements $\beta_{j,t}$ to be very persistent and excludes the case of frequent, noisy jumps. Such prior is used widely in empirical macroeconomic applications, see for example the ``business as usual'' prior motivated in \cite{CogleySargent2005} for the case of a vector autoregression with time-varying parameters. We subsequently set an uninformative prior on $\tau_{j,t}^{2}$ by setting $g_{0}=h_{0}=0.01$. The dashed lines in \autoref{fig:EMP1} represent (posterior mean) coefficient estimates from our eight-predictor model: coefficient $\beta_{1,t}$ is the time-varying intercept (trend inflation), coefficients $\beta_{2,t},\beta_{3,t}$ correspond to the first two lagged values of GDP deflator, and coefficients $\beta_{4,t}$ to $\beta_{8,t}$ correspond to the five principal components. As a comparison, we plot posterior mean estimates from the same time-varying parameter regression estimated with MCMC (using identical settings as in the Monte Carlo comparison). The MCMC-based estimates can be broadly thought of as the unrestricted equivalents of the VBDVS algorithm, since they are not based on any form of dynamic variable selection or hierarchical shrinkage. The intercept and first lag coefficients are virtually identical using the two algorithms. However, all remaining coefficients are penalized heavily by the VBDVS algorithm. Variation over time of these coefficients is very moderate and restricted to be close to zero for many time periods.

\begin{figure}[H]  
\centering
\includegraphics[scale=0.35,trim={3cm 1cm 3cm 1cm}]{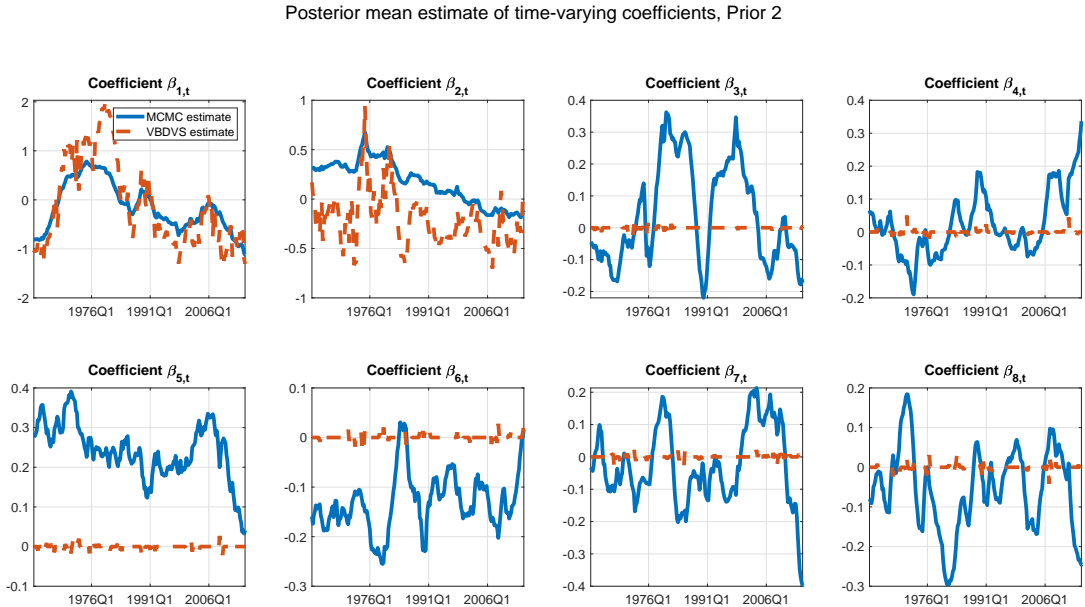}
\caption{\textit{Posterior means of time-varying coefficient estimates from VBDVS (red dashed lines) using Prior 2. Solid lines are posterior means from a TVP model with the same predictors estimated with MCMC.}}  \label{fig:EMP2}
\end{figure}

In order to examine the effect that the prior has on the time evolution of the coefficients, we change the initial condition for $w_{j,t}$ to have hyperparameters $c_{j,0}=d_{j,0}$ and we leave the same uninformative prior for $\tau_{j,t}^{2}$. The posterior mean coefficient estimates in \autoref{fig:EMP2} exhibit an interesting pattern. By allowing a looser prior on $w_{t}$ the parameters that are unrestricted (intercept and first lag), do exhibit larger amount of time-variation compared to the MCMC estimates. However, the remaining coefficients that were previously restricted to be close to zero, are now forced more aggressively towards zero in all time periods. This demonstrates the fact that our algorithm imposes the state-space model in equation \eqref{new_ss}, where the variance of $\beta_{j,t}$ is a function of both $w_{j,t}$ and $v_{j,t}$ (where the latter, is in turn a linear function of $\tau_{j,t}^{2}$). Therefore, allowing for a looser $w_{j,t}$ tends to introduce more noise in the state-space model, and for that reason the dynamic variable selection prior compensates for this increased noise by shrinking more aggressively. While there is this compensation effect and coefficient estimates won't explode as quickly as the model without the dynamic variable selection prior (recall that $\beta_{j,t}$ evolves as a non-stationary random walk), it is not advisable to use such a lose prior on $w_{j,t}$.

\begin{figure}[H]  
\centering
\includegraphics[scale=0.35,trim={3cm 1cm 3cm 1cm}]{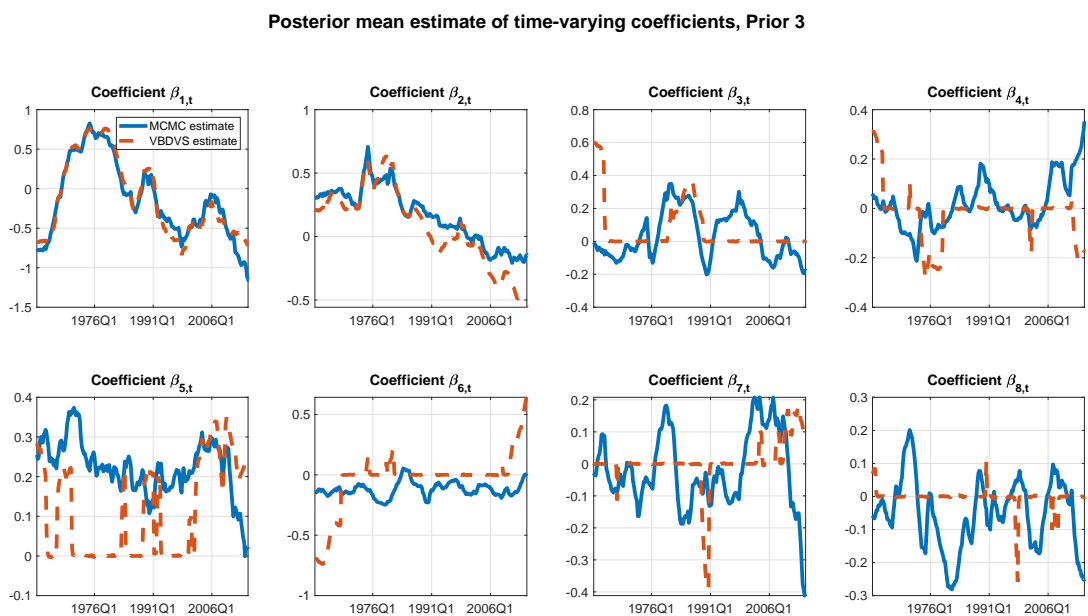}
\caption{\textit{Posterior means of time-varying coefficient estimates from VBDVS (red dashed lines) using Prior 3. Solid lines are posterior means from a TVP model with the same predictors estimated with MCMC.}}  \label{fig:EMP3}
\end{figure}

For that reason, our final prior (called Prior 3 in \autoref{table:priors}) returns to the conservative choice $c_{j,0}=100$ and $d_{j,0}=1$, and sets instead $g_{0}=1$ and $h_{0}=12$. \autoref{fig:EMP3} shows the estimates from this prior. Once again the VBDVS estimates of the intercept and first lag coefficients are identical to the estimates from the MCMC algorithm. The remaining coefficients are again heavily penalized but there are also many time periods where these evolve unrestrictedly. As a matter of fact, this prior allows the time-varying coefficients to exhibit distinct and abrupt jumps between periods where they are zero and periods where they are unrestricted. This pattern of time-variation is more in line with the findings of the previous literature that there are pockets of predictability or, put differently, that economic predictors are short-lived (see discussion in the Introduction). 

\begin{figure}[H]  
\centering
\includegraphics[scale=0.35,trim={3cm 1cm 3cm 1cm}]{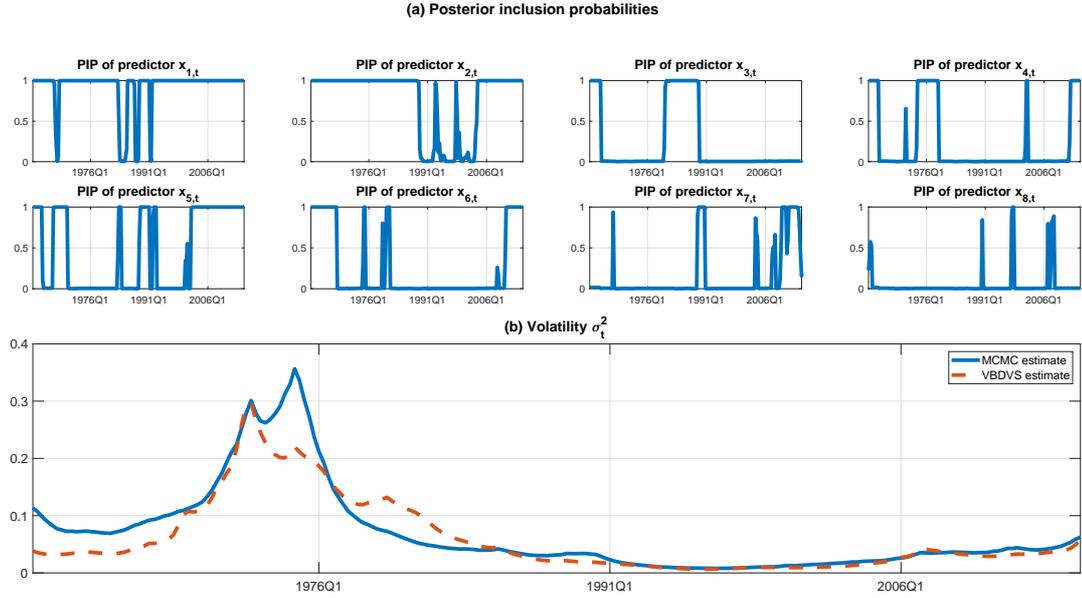}
\caption{\textit{Panel (a) shows time-varying posterior inclusion probabilities (PIPs) from VBDVS algorithm, using Prior 3. Panel (b) shows posterior means of time-varying volatility estimates from VBDVS (red dashed line) versus MCMC (solid blue line).}}  \label{fig:EMP4}
\end{figure}

In order to have a visual assessment of the time pattern of dynamic variable selection and shrinkage, panel (a) of \autoref{fig:EMP4} plots the posterior inclusion probabilities of each regressor associated with the time-varying coefficient estimates presented in \autoref{fig:EMP3}. These seem to show the exact periods where each coefficient moves from a state of being restricted to zero to a state where it is not zero. Panel (b) of the same figure shows the posterior mean of the stochastic volatility estimate from VBDVS versus the estimate from MCMC. These two estimates are fairly similar, showing that the specification of time-varying variances in the VBDVS does a good job at capturing known peaks in GDP deflator inflation volatility. Any differences in volatility estimates reflect the fact that the two algorithms assume different specification of $\sigma_{t}^{2}$ and also use different priors in the estimation of $\bm \beta_{t}$.

For all these reason, we build all of our forecasting models in the next subsection based on this last prior.\footnote{Due to the fact that the choice $h_{0}=12$ looks in \autoref{fig:EMP3} to penalize possibly excessively the small model with just eight coefficients, in the next subsection we adapt only this hyperparameter depending on the number of predictors we have available. Otherwise, all other hyperparameters are identical to the ones in the column labelled Prior 3 in \autoref{table:priors}.}

\subsection{Forecasting inflation}
We forecast inflation using models of the form
\begin{equation}
y_{t+h} = \alpha_{t} + \phi_{1,t} y_{t} + \phi_{2,t} y_{t-1} + \bm x_{t} \bm \beta_{t}  + \varepsilon_{t+h},   \label{forecasting_equation}
\end{equation}
where $y_{t+h} $ is $h$-step ahead inflation (see \autoref{sec:data} for a definition) regressed on an intercept, two own lags and exogenous predictors. We use a variety of forecasting models. Some benchmark models are based on equation \eqref{forecasting_equation} but assume constant coefficients (i.e. $\alpha_{t}=\alpha$, $\phi_{1,t} = \phi_{1}$ and so on), while others assume different sets of exogenous predictors. However, what all models have in common is that they always include an intercept and two own lags of inflation. Given that our dataset is much larger than datasets used before for forecasting inflation, in order to avoid confusion by specifying different combinations or subsets of predictors, we only distinguish four simple categories of models: i) models with no predictors (i.e. only intercept and autoregressive terms); ii) models with first five principal components as predictors; iii) models with sixty principal components as predictors; and iv) models with all 443 predictors. Our list of models representing each category is the following
\begin{itemize}
\item \textbf{AR:} benchmark AR(2) with intercept, estimated with OLS
\item \textbf{TVPAR:} time-varying parameter version of the AR model, with stochastic volatility, estimated with MCMC
\item \textbf{FAC5:} Builds on benchmark AR specification by augmenting it with first five principal components estimated with OLS
\item \textbf{BAG/FAC5:} Same predictors as FAC5, estimated as constant parameter regression using the Bagging algorithm of \cite{Breiman1996}
\item \textbf{DMA/FAC5:} Same predictors as FAC5, estimated as TVP regression using the Dynamic Model Averaging algorithm of \cite{KoopKorobilis2012}
\item \textbf{VBDVS/FAC5:} Same predictors as FAC5, estimated as TVP regression using our Dynamic Variable Selection prior with Variational Bayes 
\item \textbf{GPR/FAC5:} Same predictors as FAC5, estimated as a Gaussian Process Regression
\item \textbf{SSVS/FAC60:} Builds on benchmark AR specification by augmenting it with first 60 principal components, estimated using the SSVS prior with MCMC of \cite{GeorgeMcCulloch1993} 
\item \textbf{ELN/FAC60:} Same predictors as SSVS/FAC60, estimated as a constant parameter regression using the Elastic Net algorithm of \cite{ZouHastie2005}
\item \textbf{VBDVS/FAC60:} Same predictors as SSVS/FAC60, estimated as a TVP regression using our Dynamic Variable Selection prior with Variational Bayes
\item \textbf{ELN/X:} Builds on benchmark AR specification by augmenting it with all 443 predictors, estimated using the Elastic Net algorithm of \cite{ZouHastie2005}
\item \textbf{PLS/X:} Same predictors as in ELN/X, estimated as a constant parameter Partial Least Squares regression
\item \textbf{VBDVS/X:} Same predictors as ELN/X, estimated as a TVP regression using our Dynamic Variable Selection prior with Variational Bayes
\end{itemize}
The choice of models is based on their simplicity and replicability. In particular, the Gaussian Process Regression, Partial Least Squares, and Elastic Net algorithms are based on built-in functions in MATLAB's Statistics and Machine Learning Toolbox \citep{MatlabSTB}, and are fairly easy to set up. Estimation of these models is done using default settings in MATLAB or default choices proposed by their respective creators.\footnote{As an example, the penalty parameter in the Elastic Net is estimated using 10-fold cross-validation.} Exact details of these algorithms and their default settings is provided in the Online Supplement. 

In terms of statistical properties, all these models cover a wide spectrum of forecasting specifications. The AR(2) is a standard benchmark in economic time series forecasting, and typically performs better than a random walk (which is the benchmark for financial data). Its time-varying parameter counterpart, our second model on the list, allows for proxying for similar specifications that have been shown to forecast inflation well, see \cite{StockWatson2007} and \cite{Bauwensetal2015}. Extracting the first few principal components (factors) is possibly the most popular way of representing parsimoniously the information in a large dataset, see \cite{StockWatson2016}. A naive factor model uses least squares estimation on a model that has the first five principal components as exogenous predictors, while a second factor model replaces OLS with the Bagging algorithm of \cite{Breiman1996} that allows to select the ``best'' factors in a static way. Next the Dynamic Model Averaging (DMA) algorithm described in \cite{KoopKorobilis2012} as well as our VBDVS algorithm allow to implement dynamic variable selection in a TVP setting using the same first five principal components. The Gaussian Process Regression is a very flexible nonparametric method that allows us to understand whether inflation is better described by time-varying parameters or some more complex form of nonlinearity. Moving on to models with 60 factors, we have to drop many previous specifications for computational reasons.\footnote{For example, DMA cannot scale up to these large dimensions, Gaussian Process Regression becomes overparametrized, and Bagging becomes numerically unstable in some periods of the forecasting exercise.} For that reason we use the SSVS algorithm of \cite{GeorgeMcCulloch1993}, which can be thought of as the static equivalent of our VBDVS algorithm. The Elastic Net of \cite{ZouHastie2005} is a popular penalized likelihood estimator for high-dimensional data. Finally, our VBDVS algorithm is also estimated with a larger number of factors to find out whether its dynamic shrinkage properties are useful relative to the naive selection of the first five factors. Finally, we estimate models using all 443 exogenous predictors. The Elastic Net is again on the list, and we also include Partial Least Squares (PLS) regression. PLS is similar to principal component analysis, with the main difference being that factors are extracted with reference to the variable to be predicted. Principal components instead only explain the variability in the exogenous predictors, and it may be the case that they do not carry predictive information for the predicted variable. Finally, our VBDVS algorithm is applied to this full model with all predictors.

In terms of the prior choices used when forecasting with our VBDVS algorithm, these are based on Prior 3 described in the previous subsection, see \autoref{table:priors}. We only adapt how ``aggressively'' we shrink based on the total number of predictors in each model. For model VBDVS/FAC5 we set $h_{0}=1$, for VBDVS/FAC60 we set $h_{0}=12$ and for VBDVS/X we set $h_{0}=100$.

\begin{table}[H]
\begin{center}
\caption{\textit{Forecasting results for GDP deflator (GDPCTPI)}}  \label{table:GDPdefl_results}
\resizebox{0.9\textwidth}{!}{
\begin{tabular}{lrrrrrrrrr} \hline
	&		&	{\large \textbf{MSFE}}	&		&		&		&		&	{\large \textbf{ALPL}}	&		&		\\ \hline
	&	$h=1$	&	$h=2$	&	$h=4$	&	$h=8$	&		&	$h=1$	&	$h=2$	&	$h=4$	&	$h=8$	\\ \hline\hline
\multicolumn{10}{c}{\textsc{Models with no predictors}}																			\\
AR	&	\emph{0.0394}	&	\emph{0.0323}	&	\emph{0.0308}	&	\emph{0.0487}	&		&	\emph{4.8742}  &	\emph{4.8949}	&	\emph{4.8374}	&	\emph{4.6149}	\\
TVPAR	&	1.04	&	0.96	&	0.99	&	0.83	&		&	0.30	&	0.29	&	0.46	&	0.31	\\
\multicolumn{10}{c}{\textsc{Models with five factors}}																			\\
FAC5	&	1.00	&	1.08	&	1.53	&	1.57	&		&	0.02	&	0.04	&	0.02	&	0.01	\\
BAG/FAC5	&	0.96	&	1.05	&	1.47	&	1.48	&		&	0.05	&	0.06	&	0.04	&	0.02	\\
DMA/FAC5	&	\textbf{0.84}	&	\textbf{0.79}	&	0.94	&	0.95	&		&	0.26	&	0.27	&	0.22	&	0.18	\\
VBDVS/FAC5	&	1.30	&	1.20	&	0.97	&	0.83	&		&	0.15	&	0.14	&	0.28	&	0.20	\\
GPR/FAC5	&	1.02	&	0.95	&	1.07	&	1.04	&		&	0.06	&	0.13	&	0.20	&	0.32	\\
\multicolumn{10}{c}{\textsc{Models with 60 factors}}																			\\
SSVS/FAC60	&	0.99	&	1.03	&	1.44	&	1.42	&		&	0.01	&	0.05	&	0.09	&	0.13	\\
ELN/FAC60	&	1.13	&	1.12	&	1.24	&	1.30	&		&	0.02	&	0.07	&	0.14	&	0.09	\\
VBDVS/FAC60	&	1.03	&	0.81	&	0.85	&	0.80	&		&	0.25	&	\textbf{0.49}	&	0.63	&	\textbf{0.92}	\\
\multicolumn{10}{c}{\textsc{Models with 443 predictors}}																			\\
ELN/X	&	0.97	&	1.00	&	1.35	&	1.39	&		&	0.06	&	0.04	&	0.12	&	0.05	\\
PLS/X	&	1.14	&	1.11	&	1.42	&	1.24	&		&	-0.11	&	0.02	&	-0.24	&	-0.42	\\
VBDVS/X	&	0.99	&	0.84	&	\textbf{0.71}	&	\textbf{0.62}	&	{\hspace{2em}}	&	\textbf{0.32}	&	0.39	&	\textbf{0.65}	&	0.78	\\   \hline
\end{tabular}
}
\end{center}
{\emph{\scriptsize Notes: All models feature an intercept and two lags of the dependent variable. Model acronyms are as follows:\\
\textbf{AR:} benchmark AR(2) with intercept estimated with OLS \\
\textbf{TVPAR:} time-varying parameter version of the AR model, with stochastic volatility, estimated with MCMC \\
\textbf{FAC5:} Builds on benchmark AR specification by augmenting it with first five principal components estimated with OLS \\
\textbf{BAG/FAC5:} Same predictors as FAC5, estimated as constant parameter regression using Bagging \\
\textbf{DMA/FAC5:} Same predictors as FAC5, estimated as TVP regression using Dynamic Model Averaging \\
\textbf{VBDVS/FAC5:} Same predictors as FAC5, estimated as TVP regression using our Dynamic Variable Selection prior with Variational Bayes \\
\textbf{GPR/FAC5:} Same predictors as FAC5, estimated as a Gaussian Process Regression \\
\textbf{SSVS/FAC60:} Builds on benchmark AR specification by augmenting it with first 60 principal components, estimated using an SSVS prior with MCMC \\
\textbf{ELN/FAC60:} Same predictors as SSVS/FAC60, estimated as a constant parameter regression using the Elastic Net \\
\textbf{VBDVS/FAC60:} Same predictors as SSVS/FAC60, estimated as a TVP regression using our Dynamic Variable Selection prior with Variational Bayes \\
\textbf{ELN/X:} Builds on benchmark AR specification by augmenting it with all 443 predictors, estimated using the Elastic Net \\
\textbf{PLS/X:} Same predictors as in ELN/X, estimated as a constant parameter Partial Least Squares regression \\
\textbf{VBDVS/X:} Same predictors as ELN/X, estimated as a TVP regression using our Dynamic Variable Selection prior with Variational Bayes \\
Entries in columns 2-5 of this Table are mean squared forecast errors (MSFEs), and columns 6-9 are average predictive likelihoods in logarithms (logAPLs). The AR model serves as a benchmark and its entries (shown in italics) are the values of MSFEs and logAPLs for each forecast horizon. Entries for each subsequent model are MSFEs and logAPLs \underline{relative} to the values of the AR benchmark. MSFEs lower than one signify improvement relative to the benchmark and vice-versa for values higher than one. logAPLs that are positive signify improvement relative to the benchmark and vice-versa for negative values. Entries in boldface indicate the best performing model for each forecast statistic and for each forecast horizon.} }
\end{table}

\begin{table}[H]
\begin{center}
\caption{\textit{Forecasting results for PCE deflator (PCECTPI)}}  \label{table:PCEdefl_results}
\resizebox{0.9\textwidth}{!}{
\begin{tabular}{lrrrrrrrrr} \hline
	&		&	{\large \textbf{MSFE}}	&		&		&		&		&	{\large \textbf{ALPL}}	&		&		\\ \hline
	&	$h=1$	&	$h=2$	&	$h=4$	&	$h=8$	&		&	$h=1$	&	$h=2$	&	$h=4$	&	$h=8$	\\ \hline\hline
\multicolumn{10}{c}{\textsc{Models with no predictors}}																			\\
AR	&	\emph{0.1442}	&	\emph{0.1301}	&	\emph{0.1070}	&	\emph{0.0980}	&		&	\emph{4.6028}	&	\emph{4.6527}	&	\emph{4.5695}	&	\emph{4.4313}	\\
TVPAR	&	1.10	&	1.04	&	0.81	&	0.57	&		&	0.08	&	0.24	&	\textbf{0.62}	&	0.67	\\
\multicolumn{10}{c}{\textsc{Models with five factors}}																			\\
FAC5	&	1.10	&	1.23	&	1.32	&	1.40	&		&	0.02	&	0.03	&	0.05	&	0.06	\\
BAG/FAC5	&	1.13	&	1.27	&	1.33	&	1.38	&		&	0.04	&	0.06	&	0.07	&	0.05	\\
DMA/FAC5	&	1.14	&	1.13	&	1.02	&	0.86	&		&	-0.07	&	-0.02	&	0.05	&	0.27	\\
VBDVS/FAC5	&	0.92	&	0.95	&	0.75	&	0.71	&		&	-0.09	&	-0.14	&	-0.21	&	0.29	\\
GPR/FAC5	&	1.08	&	1.17	&	1.00	&	0.96	&		&	0.06	&	0.22	&	0.31	&	0.23	\\
\multicolumn{10}{c}{\textsc{Models with 60 factors}}																			\\
SSVS/FAC60	&	0.98	&	1.18	&	1.19	&	1.37	&		&	0.07	&	0.11	&	0.17	&	0.28	\\
ELN/FAC60	&	0.84	&	1.04	&	1.03	&	0.95	&		&	0.16	&	0.20	&	0.24	&	0.32	\\
VBDVS/FAC60	&	1.11	&	1.04	&	0.70	&	0.51	&		&	0.05	&	0.22	&	0.54	&	\textbf{1.02}	\\
\multicolumn{10}{c}{\textsc{Models with 443 predictors}}																			\\
ELN/X	&	\textbf{0.73}	&	0.97	&	1.06	&	1.00	&		&	\textbf{0.22}	&	0.13	&	0.12	&	0.02	\\
PLS/X	&	0.81	&	0.91	&	0.95	&	0.82	&		&	0.12	&	0.07	&	0.15	&	-0.05	\\
VBDVS/X	&	0.93	&	\textbf{0.84}	&	\textbf{0.62}	&	\textbf{0.51}	&	{\hspace{2em}}	&	0.11	&	\textbf{0.27}	&	0.57	&	0.69	\\
 \hline
\end{tabular}
}
\end{center}
{\emph{\scriptsize Notes: see notes under \autoref{table:GDPdefl_results}. }}
\end{table}

\begin{table}[H]
\begin{center}
\caption{\textit{Forecasting results for CPI (CPIAUCSL)}}  \label{table:CPI_results}
\resizebox{0.9\textwidth}{!}{
\begin{tabular}{lrrrrrrrrr} \hline
	&		&	{\large \textbf{MSFE}}	&		&		&		&		&	{\large \textbf{ALPL}}	&		&		\\ \hline
	&	$h=1$	&	$h=2$	&	$h=4$	&	$h=8$	&		&	$h=1$	&	$h=2$	&	$h=4$	&	$h=8$	\\ \hline\hline
\multicolumn{10}{c}{\textsc{Models with no predictors}}																			\\
AR	&	\emph{0.1838}	&	\emph{0.2247}	&	\emph{0.1621}	&	\emph{0.1425}	&		&	\emph{4.3741}	&	\emph{4.3776}	&	\emph{4.3663}	&	\emph{4.2359}	\\
TVPAR	&	0.93	&	0.96	&	0.73	&	0.57	&		&	0.09	&	0.42	&	\textbf{0.59}	&	0.96	\\
\multicolumn{10}{c}{\textsc{Models with five factors}}																			\\
FAC5	&	0.95	&	0.96	&	1.05	&	1.03	&		&	0.04	&	0.09	&	0.03	&	0.16	\\
BAG/FAC5	&	0.95	&	0.97	&	1.01	&	0.92	&		&	0.06	&	0.10	&	0.09	&	0.16	\\
DMA/FAC5	&	0.93	&	0.91	&	0.87	&	0.67	&		&	-0.02	&	0.00	&	0.02	&	0.35	\\
VBDVS/FAC5	&	1.07	&	1.15	&	0.71	&	0.75	&		&	0.06	&	0.43	&	-0.12	&	0.40	\\
GPR/FAC5	&	0.93	&	0.90	&	0.79	&	0.82	&		&	0.12	&	0.30	&	0.29	&	0.25	\\
\multicolumn{10}{c}{\textsc{Models with 60 factors}}																			\\
SSVS/FAC60	&	0.89	&	0.87	&	0.93	&	0.87	&		&	0.06	&	0.13	&	0.14	&	0.31	\\
ELN/FAC60	&	1.08	&	\textbf{0.81}	&	0.89	&	0.72	&		&	-0.01	&	0.20	&	0.22	&	0.39	\\
VBDVS/FAC60	&	0.98	&	0.88	&	0.67	&	0.54	&		&	0.10	&	0.45	&	\textbf{0.56}	&	\textbf{1.01}	\\
\multicolumn{10}{c}{\textsc{Models with 443 predictors}}																			\\
ELN/X	&	\textbf{0.84}	&	0.84	&	0.92	&	0.89	&		&	0.21	&	0.27	&	0.19	&	0.23	\\
PLS/X	&	0.95	&	0.89	&	0.96	&	0.89	&		&	0.08	&	0.34	&	0.21	&	0.26	\\
VBDVS/X	&	0.94	&	0.82	&	\textbf{0.64}	&	\textbf{0.50}	&	{\hspace{2em}}	&	\textbf{0.24}	&	0.33	&	0.47	&	0.81	\\
 \hline
\end{tabular}
}
\end{center}
{\emph{\scriptsize Notes: see notes under \autoref{table:GDPdefl_results}. }}
\end{table}

\begin{table}[H]
\begin{center}
\caption{\textit{Forecasting results for core CPI (CPILFESL)}}  \label{table:coreCPI_results}
\resizebox{0.9\textwidth}{!}{
\begin{tabular}{lrrrrrrrrr} \hline
	&		&	{\large \textbf{MSFE}}	&		&		&		&		&	{\large \textbf{ALPL}}	&		&		\\ \hline
	&	$h=1$	&	$h=2$	&	$h=4$	&	$h=8$	&		&	$h=1$	&	$h=2$	&	$h=4$	&	$h=8$	\\ \hline\hline
\multicolumn{10}{c}{\textsc{Models with no predictors}}																			\\
AR	&	\emph{0.0221}	&	\emph{0.0195}	&	\emph{0.0287}	&	\emph{0.0549}	&		&	\emph{4.6432}	&	\emph{4.6832}	&	\emph{4.6294}	&	\emph{4.4824}	\\
TVPAR	&	1.00	&	0.85	&	0.79	&	0.48	&		&	0.52	&	0.55	&	0.58	&	0.54	\\
\multicolumn{10}{c}{\textsc{Models with five factors}}																			\\
FAC5	&	1.89	&	2.37	&	2.17	&	1.41	&		&	0.02	&	0.04	&	0.09	&	0.15	\\
BAG/FAC5	&	1.62	&	2.15	&	1.90	&	1.26	&		&	0.06	&	0.08	&	0.14	&	0.16	\\
DMA/FAC5	&	1.17	&	1.23	&	0.94	&	0.60	&		&	0.36	&	0.42	&	0.48	&	0.56	\\
VBDVS/FAC5	&	1.49	&	1.24	&	0.90	&	0.53	&		&	0.54	&	0.39	&	0.67	&	0.48	\\
GPR/FAC5	&	1.66	&	1.79	&	1.44	&	0.99	&		&	0.34	&	0.37	&	0.68	&	0.50	\\
\multicolumn{10}{c}{\textsc{Models with 60 factors}}																			\\
SSVS/FAC60	&	1.73	&	2.18	&	2.00	&	1.05	&		&	0.03	&	0.04	&	0.13	&	0.19	\\
ELN/FAC60	&	1.91	&	2.09	&	1.81	&	0.99	&		&	0.05	&	0.04	&	0.26	&	0.32	\\
VBDVS/FAC60	&	\textbf{0.91}	&	0.79	&	0.72	&	0.47	&		&	0.72	&	\textbf{0.78}	&	0.93	&	0.92	\\
\multicolumn{10}{c}{\textsc{Models with 443 predictors}}																			\\
ELN/X	&	1.79	&	1.96	&	1.47	&	1.21	&		&	0.22	&	0.11	&	0.37	&	-0.05	\\
PLS/X	&	2.53	&	2.87	&	2.05	&	1.27	&		&	0.13	&	0.00	&	0.14	&	0.19	\\
VBDVS/X	&	0.99	&	\textbf{0.78}	&	\textbf{0.60}	&	\textbf{0.43}	&	{\hspace{2em}}	&	\textbf{0.71}	&	\textbf{0.78}	&	\textbf{0.99}	&	\textbf{1.07}	\\
 \hline
\end{tabular}
}
\end{center}
{\emph{\scriptsize Notes: see notes under \autoref{table:GDPdefl_results}. }}
\end{table}

We forecast $h=1,2,4$ and $8$ quarters ahead. We use 50\% of the sample as our initial estimation period which, for example, for $h=1$ translates to using data for the period 1960Q4-1989Q2 in order to forecast 1989Q3. We then add one new observation to the estimation sample and forecast $h$-step ahead, until the full sample is exhausted. Since all models that have predictors rely on the direct forecasting regression \eqref{forecasting_equation}, for comparability we produce direct AR(2) forecasts as a special case of this equation with no predictors.\footnote{The alternative would be to specify an AR(2) model linking $y_{t}$ with $y_{t-1}$ and $y_{t-2}$ and then iterate the process $h$ periods ahead, a procedure also known as iterative forecasting. By using direct AR(2) forecasts as the benchmark we can explicitly assess the exact contribution of various models that introduce exogenous predictors.} We measure forecast accuracy using the mean squared forecast error (MSFE) and the average log-predictive likelihood (ALPL). The first measure is the square of the forecast error (difference between forecast and real value of $y_{t+h}$) averaged over the out-of-sample evaluation period, while the second measure is calculated as the logarithm of the predictive distribution evaluated at the observation $y_{t+h}$ and also averaged over the out-of-sample evaluation period; see \cite{Bauwensetal2015} for more details on these two metrics.

\cref{table:GDPdefl_results,table:PCEdefl_results,table:CPI_results,table:coreCPI_results} present the MSFEs and ALPLs for GDP deflator, PCE deflator, CPI and Core CPI, for all competing models and all considered forecast horizons. To be precise results for the benchmark AR(2) are the values of the MSFE and ALPL statistics, while results for all other models are relative to those for the AR(2). For the MSFE this means calculating the ratio such that a number lower than one means that a certain model performs better than the AR(2). For the ALPL relative quantities are obtained as the spread from the ALPL of the AR(2) (i.e. the logarithm of the ratio) such that positive numbers indicate that a certain model performs better than the AR(2).

The immediate message from these tables is that the VBDVS/X is the model that performs best, especially when looking at point forecast evaluation (MSFEs) for $h=2,4,8$. In terms of density forecasts, VBDVS/X and VBDVS/FAC60 are jointly the best performing specifications. While VBDVS/FAC5 is also doing well in longer horizons, this model is always underperforming the TVPAR, that is, the TVP model that doesn't consider any predictors.

How can we explain these results? There are various stylized facts we can derive from the information in these tables. Our discussion here focuses on point forecasts (MSFE criterion), due to the fact for that metric the picture is much clearer. First, time variation seems to matter a lot, especially in the long-run. TVPAR, DMA/FAC5, and the three VBDVS specifications can improve dramatically over their constant parameter counterparts, regardless of whether these consider exogenous predictors or not. Are exogenous predictors important for forecasting? The answer depends on the variable to be forecast, the horizon considered, as well as the way each model specification utilizes the predictors. For example, for GDP deflator for $h=8$ the differences in MSFE between VBDVS/X (TVP model with all available predictors) and TVPAR (TVP model with no predictors) is vast, suggesting that not only time-variation is important but also the information in exogenous predictors. However, looking at all constant parameter models with exogenous predictors, whether these predictors are observed or enter each regression via factor methods, all these methods struggle to beat the simple AR(2). This suggests the argument in the Introduction about pockets of predictability. For that reason, DMA (which is the best performing model for $h=1$ and $h=2$) and the three VBDVS specifications perform very well, with VBDVS/X providing the most dramatic improvements for $h=8$ when at the same time ELN/X and PLS/X perform 24\% and 39\% worse than the benchmark AR(2).

For the next two inflation variables (PCE deflator and total CPI) a large number of predictors does seem to be important in the short-run, but in the long-run it looks like the largest contribution in forecasting accuracy is due to time-variation in parameters. For example, for PCE deflator and total CPI, for horizons $h=1,2$, ELN/X seems to be performing much better than the AR and the TVPAR specifications. However, for $h=4,8$ the TVPAR overtakes substantially both the AR and ELN/X specifications. While the VBDVS/X is still the best performing model for $h=4,8$, its differences to the TVPAR are statistically much smaller compared to the differences of these two models when forecasting GDP deflator. In any case, whether predictors are important or not, the VBDVS algorithm seems to be doing a very good job in shrinking irrelevant coefficients and making sure that there is not overfitting -- if there was, the VBDVS/X forecasts would be inferior to those from the TVPAR.

Finally, for core CPI all methods struggle to beat the simple AR for very short-run forecasts. The VBDVS/FAC60 and VBDVS/X models do so marginally, while many others perform as much as 150\% worse than the benchmark. For longer horizons all constant parameter models continue to underperform, however, the TVP models seem to provide the most dramatic improvements, with the VBDVS/X improving almost 60\% over the benchmark. Combined with the observation that the differences between the three VBDVS specifications and the TVPAR are minimal, it looks like that exogenous predictors are not relevant for core CPI. Since core CPI is based on the total CPI by removing its most volatiles components (food and energy), it might be the case that this variable is basically a random walk and even a simple time-varying intercept model (that is, a local level model as in \citealp{StockWatson2007}) would forecast this variable well.

It is harder to extract stylized facts for inflation forecasting based on ALPLs. This is because this metric is based on all the features of the predictive density, that is, all its moments and not just the mean. Given that predictive densities can differ a lot between specifications (e.g. they can multimodal in time-varying parameter models), it is not possible to attribute differences in ALPLs to specific modeling assumptions. However, a clear pattern that emerges is that predictors do help to improve predictive density forecasting relative to the simple AR benchmark, but the largest gains overall are achieved by time-varying parameter models. In all these comparisons the VBDVS/X is the clear winner showing that, even though this is a heavily parametrized model and could easily produce erroneous forecasts, our algorithm ensures sufficient penalization and impressive forecasting gains.

\newpage
\addcontentsline{toc}{section}{\refname}
\bibliographystyle{ecta}
\bibliography{VBKF_revisions}

\newpage
\singlespace
\setcounter{page}{1}
\begin{appendix}
\begin{center}
{\LARGE Online Supplement to ``Bayesian dynamic variable selection in high-dimensions'' } \\
\hfill \\
{\large         Gary Koop   \hspace{2cm}   Dimitris Korobilis}
\end{center}
\section{Settings used in competing models}
While all technical details regarding our methodology are provided in detail in the paper, we have skipped details for the numerous competing algorithms used in the Monte Carlo and empirical exercises.
\begin{itemize}
\item \textbf{ DSS algorithm, \cite{RockovaMcAlinn2017}}: We followed the authors and tried the various settings they suggest in their Section 7: Synthetic high-dimensional data. For our DGP the best performance was achieved with $\phi_{0}=0$, $\phi_{1}= 0.98$, $\lambda_{1}= 10*(1 - \phi_{1}.^2)$, $\lambda_{0}=0.9$ and $\Theta = 0.92$ (note that for $p=50$ the authors suggest $\Theta=0.98$, but we found that a lower value does better as $p$ gets larger, while it doesn't deteriorate performance for $p=50$).
\item \textbf{MCMC algorithm, \cite{ChanJeliazkov2009}}: This is the standard time-varying parameter regression model used in economics, see for example \cite{CogleySargent2005}. It consists of equations \eqref{TVP_measurement} and \eqref{TVP_state}, where the measurement error variance follows a geometric random walk. As with VBDVS, the crucial setting that affects the amount of time-variation in regression coefficients is the prior on the state variances, which is of the form $w_{j}^{-1} \sim Gamma(v_{1},v_{2})$. We set the conservative choice $v_{1}=3$ and $v_{2} = 20$, which implies that $w_{j}$ has prior mean around $0.016$. In order to estimate this model efficiently, we use the Gibbs sampler algorithm of \cite{ChanJeliazkov2009}.
\item \textbf{Dynamic Model Averaging, \cite{KoopKorobilis2012}:} We use standard settings described in \cite{KoopKorobilis2012} with $\alpha=0.99$, $\lambda =0.99$
and $\kappa=0.96$.
\item \textbf{Bagging, \cite{Breiman1996}:} With the bagging algorithm we first resample our data $B$ times with replacement blocks of size $m$. For each pseudo-generated dataset we estimate with ordinary least squares using the Newey and West estimator of the covariance with lag truncation parameter $int \left \lbrace T^{1/4} \right \rbrace$. We select the optimal model using only those predictors that have t-statistics larger than a threshold $c^{\ast}$ in absolute value. We forecast with the optimal model, and the bagging forecast is obtained as the average of all forecasts over the $B$ Bootstrap replications. We set $B=1000$, $m=1$ and $c^{\ast} = 2.807$.
\item \textbf{Elastic Net, \cite{ZouHastie2005}:} We use the MATLAB function ``lasso'' that is available in the Statistics and Machine Learning Toolbox. We use 10-fold cross validation for selecting the optimal $\lambda$ parameter, and we fix $\alpha=0.75$.
\item \textbf{Gaussian Process Regression:} Gaussian Process Regression (GPR) is a very powerful machine learning method that allows flexible nonparametric estimation targeted towards prediction. We use the MATLAB function ``fitrgp'' that is available in the Statistics and Machine Learning Toolbox. This is estimated using the following settings: \\
\texttt{fitrgp(X,y,'Basis','linear','Optimizer','QuasiNewton','verbose',1,} \texttt{'FitMethod','exact','PredictMethod','exact')}
\item \textbf{Partial Least Squares:} Partial Least Squares (PLS) is a method that originated in chemometrics. It allows to estimate factors that are extracted with reference to the variable to be predicted (target variable). Principal components instead maximize only the variance explained by the large dataset, and may not be optimal for prediction of the target variable. While more elegant methods have been proposed recently, such as the three-pass regression filter, the PLS is undeniably a good benchmark for assessing whether we can improve on the information content of simple principal component estimates. We use again the MATLAB function ``plsregress'' available in the Statistics and Machine Learning Toolbox, and we extract five factors from our dataset.
\end{itemize}

\newpage
\begin{landscape}
\renewcommand{\theequation}{A.\arabic{equation}} \setcounter{equation}{0} 
\renewcommand{\thetable}{A\arabic{table}} \setcounter{table}{0}
\section{Data Appendix}
The following high-dimensional dataset combines several popular datasets used in macroeconomics and finance. The core part builds on the FRED-QD dataset compiled in \cite{McCrackenNg2020}, and the financial (portfolio) data used in \cite{Juradoetal2015} to extract a popular uncertainty index that are originally provided by Kenneth French (\href{https://mba.tuck.dartmouth.edu/pages/faculty/ken.french/data_library.html}{https://mba.tuck.dartmouth.edu/pages/faculty/ken.french/data\_library.html}). These are augmented with additional consumer survey indicators from University of Michigan (\href{https://data.sca.isr.umich.edu/}{https://data.sca.isr.umich.edu/}); predictors of stock returns used in \cite{WelchGoyal2008} provided by Amit Goyal (\href{http://www.hec.unil.ch/agoyal/}{http://www.hec.unil.ch/agoyal/}); Commodity prices from the World Bank's Pink Sheet database (\href{https://www.worldbank.org/en/research/commodity-markets}{https://www.worldbank.org/en/research/commodity-markets}); and key macroeconomic indicators for key economies, obtained from Federal Reserve Economic Data (FRED) of St Louis Federal Reserve Bank (\href{https://fred.stlouisfed.org/}{https://fred.stlouisfed.org/}).

\autoref{datatable} presents the 444 variables used in the empirical exercise. These are measured quarterly and cover the period 1960Q1-2018Q4. Where a variable is measured originally in higher frequency (e.g. monthly) quarterly values are obtained by taking the average over the quarter. Column $F$ in \autoref{datatable} denotes whether the variable is used or not (1 or 0, respectively) to extract factors (principal components). The idea is that where some variables are aggregates of disaggregated series in the dataset, we only use the disaggregated series to extract factors. Column $F$ denotes the code used in order to transform each variable to be approximately stationary. The transformation codes are the following 1: level (no transformation); 2: first difference; 3: second difference; 4: natural logarithm; 5: first difference of natural logarithm; 6: second difference of natural logarithm; 7: first difference of percent change. The mnemonics used are those provided by the respective resources. For the \cite{WelchGoyal2008} data in particular, the mnemonics are those provided in the data appendix of that paper. In the $Source$ column of \autoref{datatable}, this paper is abbreviated as GW2008.

\clearpage

{\scriptsize
\begin{longtable}{llccll}
\caption[Quarterly large dataset]{Quarterly large dataset} \label{datatable} \\
No & Mnemonic	&	F	&	T	&	Long description & Source	\\ \hline \hline \endfirsthead
 & & & & & \\ \hline \endfoot
\multicolumn{6}{l}{Table \ref{datatable} (continued)} \\ \hline \hline \endhead
1	&	GDPC1	&	0	&	5	&	Real Gross Domestic Product	&	FRED-QD	\\
2	&	PCECC96	&	0	&	5	&	Consumption Real Personal Consumption Expenditures	&	FRED-QD	\\
3	&	PCDGx	&	1	&	5	&	Real personal consumption expenditures: Durable goods	&	FRED-QD	\\
4	&	PCESVx	&	1	&	5	&	Real Personal Consumption Expenditures: Services	&	FRED-QD	\\
5	&	PCNDx	&	1	&	5	&	Real Personal Consumption Expenditures: Nondurable Goods	&	FRED-QD	\\
6	&	GPDIC1	&	0	&	5	&	Real Gross Private Domestic Investment	&	FRED-QD	\\
7	&	FPIx	&	0	&	5	&	Real private fixed investment	&	FRED-QD	\\
8	&	Y033RC1Q027SBEAx	&	1	&	5	&	Real Gross Private Domestic Fixed Investment: Nonresidential: Equipment	&	FRED-QD	\\
9	&	PNFIx	&	1	&	5	&	Real private fixed investment: Nonresidential	&	FRED-QD	\\
10	&	PRFIx	&	1	&	5	&	Real private fixed investment: Residential	&	FRED-QD	\\
11	&	A014RE1Q156NBEA	&	1	&	1	&	Gross private domestic investment: Change in private inventories	&	FRED-QD	\\
12	&	GCEC1	&	0	&	5	&	Real Government Consumption Expenditures \& Gross Investment	&	FRED-QD	\\
13	&	A823RL1Q225SBEA	&	1	&	1	&	Real Government Consumption Expenditures and Gross Investment: Federal	&	FRED-QD	\\
14	&	FGRECPTx	&	1	&	5	&	Real Federal Government Current Receipts	&	FRED-QD	\\
15	&	SLCEx	&	1	&	5	&	Real government state and local consumption expenditures	&	FRED-QD	\\
16	&	EXPGSC1	&	1	&	5	&	Real Exports of Goods \& Services	&	FRED-QD	\\
17	&	IMPGSC1	&	1	&	5	&	Real Imports of Goods \& Services	&	FRED-QD	\\
18	&	DPIC96	&	0	&	5	&	Real Disposable Personal Income	&	FRED-QD	\\
19	&	OUTNFB	&	0	&	5	&	Real Disposable Personal Income	&	FRED-QD	\\
20	&	OUTBS	&	0	&	5	&	Business Sector: Real Output	&	FRED-QD	\\
21	&	INDPRO	&	0	&	5	&	Industrial Production Index	&	FRED-QD	\\
22	&	IPFINAL	&	0	&	5	&	Industrial Production: Final Products	&	FRED-QD	\\
23	&	IPCONGD	&	0	&	5	&	Industrial Production: Consumer Goods	&	FRED-QD	\\
24	&	IPMAT	&	0	&	5	&	Industrial Production: Materials	&	FRED-QD	\\
25	&	IPDMAT	&	1	&	5	&	Industrial Production: Durable Materials	&	FRED-QD	\\
26	&	IPNMAT	&	1	&	5	&	Industrial Production: Nondurable Materials	&	FRED-QD	\\
27	&	IPDCONGD	&	1	&	5	&	Industrial Production: Durable Consumer Goods	&	FRED-QD	\\
28	&	IPB51110SQ	&	1	&	5	&	Industrial Production: Durable Goods: Automotive products	&	FRED-QD	\\
29	&	IPNCONGD	&	1	&	5	&	Industrial Production: Durable Goods: Automotive products	&	FRED-QD	\\
30	&	IPBUSEQ	&	1	&	5	&	Industrial Production: Business Equipment	&	FRED-QD	\\
31	&	IPB51220SQ	&	1	&	5	&	Industrial Production: Consumer energy products	&	FRED-QD	\\
32	&	CUMFNS	&	1	&	1	&	Capacity Utilization: Manufacturing (SIC)	&	FRED-QD	\\
33	&	PAYEMS	&	0	&	5	&	All Employees: Total nonfarm	&	FRED-QD	\\
34	&	USPRIV	&	0	&	5	&	All Employees: Total Private Industries	&	FRED-QD	\\
35	&	MANEMP	&	0	&	5	&	All Employees: Manufacturing	&	FRED-QD	\\
36	&	SRVPRD	&	0	&	5	&	All Employees: Service-Providing Industries	&	FRED-QD	\\
37	&	USGOOD	&	0	&	5	&	All Employees: Goods-Producing Industries	&	FRED-QD	\\
38	&	DMANEMP	&	1	&	5	&	All Employees: Durable goods	&	FRED-QD	\\
39	&	NDMANEMP	&	0	&	5	&	All Employees: Nondurable goods	&	FRED-QD	\\
40	&	USCONS	&	1	&	5	&	All Employees: Construction	&	FRED-QD	\\
41	&	USEHS	&	1	&	5	&	All Employees: Education \& Health Services	&	FRED-QD	\\
42	&	USFIRE	&	1	&	5	&	All Employees: Education \& Health Services	&	FRED-QD	\\
43	&	USINFO	&	1	&	5	&	All Employees: Information Services	&	FRED-QD	\\
44	&	USPBS	&	1	&	5	&	All Employees: Professional \& Business Services	&	FRED-QD	\\
45	&	USLAH	&	1	&	5	&	All Employees: Leisure \& Hospitality	&	FRED-QD	\\
46	&	USSERV	&	1	&	5	&	All Employees: Other Services	&	FRED-QD	\\
47	&	USMINE	&	1	&	5	&	All Employees: Mining and logging	&	FRED-QD	\\
48	&	USTPU	&	1	&	5	&	All Employees: Trade, Transportation \& Utilities	&	FRED-QD	\\
49	&	USGOVT	&	0	&	5	&	All Employees: Government	&	FRED-QD	\\
50	&	USTRADE	&	1	&	5	&	All Employees: Retail Trade	&	FRED-QD	\\
51	&	USWTRADE	&	1	&	5	&	All Employees: Wholesale Trade	&	FRED-QD	\\
52	&	CES9091000001	&	1	&	5	&	All Employees: Government: Federal	&	FRED-QD	\\
53	&	CES9092000001	&	1	&	5	&	All Employees: Government: State Government	&	FRED-QD	\\
54	&	CES9093000001	&	1	&	5	&	All Employees: Government: Local Government	&	FRED-QD	\\
55	&	CE16OV	&	0	&	5	&	Civilian Employment	&	FRED-QD	\\
56	&	CIVPART	&	0	&	2	&	Civilian Labor Force Participation Rate	&	FRED-QD	\\
57	&	UNRATE	&	0	&	2	&	Civilian Unemployment Rate	&	FRED-QD	\\
58	&	UNRATESTx	&	0	&	2	&	Unemployment Rate less than 27 weeks	&	FRED-QD	\\
59	&	UNRATELTx	&	0	&	2	&	Unemployment Rate for more than 27 weeks	&	FRED-QD	\\
60	&	LNS14000012	&	1	&	2	&	Unemployment Rate - 16 to 19 years	&	FRED-QD	\\
61	&	LNS14000025	&	1	&	2	&	Unemployment Rate - 20 years and over, Men	&	FRED-QD	\\
62	&	LNS14000026	&	1	&	2	&	Unemployment Rate - 20 years and over, Women	&	FRED-QD	\\
63	&	UEMPLT5	&	1	&	5	&	Number of Civilians Unemployed - Less Than 5 Weeks	&	FRED-QD	\\
64	&	UEMP5TO14	&	1	&	5	&	Number of Civilians Unemployed for 5 to 14 Weeks	&	FRED-QD	\\
65	&	UEMP15T26	&	1	&	5	&	Number of Civilians Unemployed for 15 to 26 Weeks	&	FRED-QD	\\
66	&	UEMP27OV	&	1	&	5	&	Number of Civilians Unemployed for 27 Weeks and Over	&	FRED-QD	\\
67	&	LNS12032194	&	1	&	5	&	Employment Level - Part-Time for Economic Reasons, All Industries	&	FRED-QD	\\
68	&	HOABS	&	0	&	5	&	Business Sector: Hours of All Persons	&	FRED-QD	\\
69	&	HOANBS	&	0	&	5	&	Nonfarm Business Sector: Hours of All Persons	&	FRED-QD	\\
70	&	AWHMAN	&	1	&	1	&	Average Weekly Hours of Production and Nonsupervisory Employees: Manufacturing	&	FRED-QD	\\
71	&	AWOTMAN	&	1	&	2	&	Average Weekly Hours Of Production And Nonsupervisory Employees: Total private	&	FRED-QD	\\
72	&	HWIx	&	0	&	1	&	Help-Wanted Index	&	FRED-QD	\\
73	&	HOUST	&	0	&	5	&	Housing Starts: Total: New Privately Owned Housing Units Started	&	FRED-QD	\\
74	&	HOUST5F	&	0	&	5	&	Housing Starts: Total: New Privately Owned Housing Units Started	&	FRED-QD	\\
75	&	PERMIT	&	1	&	5	&	New Private Housing Units Authorized by Building Permits	&	FRED-QD	\\
76	&	HOUSTMW	&	1	&	5	&	Housing Starts in Midwest Census Region	&	FRED-QD	\\
77	&	HOUSTNE	&	1	&	5	&	Housing Starts in Northeast Census Region	&	FRED-QD	\\
78	&	HOUSTS	&	1	&	5	&	Housing Starts in South Census Region	&	FRED-QD	\\
79	&	HOUSTW	&	1	&	5	&	Housing Starts in West Census Region	&	FRED-QD	\\
80	&	CMRMTSPLx	&	0	&	5	&	Real Manufacturing and Trade Industries Sales	&	FRED-QD	\\
81	&	RSAFSx	&	1	&	5	&	Real Retail and Food Services Sales	&	FRED-QD	\\
82	&	AMDMNOx	&	1	&	5	&	Real Manufacturers’ New Orders: Durable Goods	&	FRED-QD	\\
83	&	AMDMUOx	&	1	&	5	&	Real Manufacturers’ Unfilled Orders for Durable Goods	&	FRED-QD	\\
84	&	PCECTPI	&	0	&	6	&	Personal Consumption Expenditures: Chain-type Price Index	&	FRED-QD	\\
85	&	PCEPILFE	&	0	&	6	&	Personal Consumption Expenditures Excluding Food and Energy	&	FRED-QD	\\
86	&	GDPCTPI	&	0	&	6	&	Gross Domestic Product: Chain-type Price Index	&	FRED-QD	\\
87	&	GPDICTPI	&	1	&	6	&	Gross Private Domestic Investment: Chain-type Price Index	&	FRED-QD	\\
88	&	IPDBS	&	1	&	6	&	Business Sector: Implicit Price Deflator	&	FRED-QD	\\
89	&	DGDSRG3Q086SBEA	&	0	&	6	&	Goods Personal consumption expenditures: Goods	&	FRED-QD	\\
90	&	DDURRG3Q086SBEA	&	0	&	6	&	Personal consumption expenditures: Durable goods	&	FRED-QD	\\
91	&	DSERRG3Q086SBEA	&	0	&	6	&	Personal consumption expenditures: Services	&	FRED-QD	\\
92	&	DNDGRG3Q086SBEA	&	0	&	6	&	Personal consumption expenditures: Nondurable goods 	&	FRED-QD	\\
93	&	DHCERG3Q086SBEA	&	0	&	6	&	Personal consumption expenditures: Nondurable goods 	&	FRED-QD	\\
94	&	DMOTRG3Q086SBEA	&	1	&	6	&	Personal consumption expenditures: Durable goods: Motor vehicles and parts	&	FRED-QD	\\
95	&	DFDHRG3Q086SBEA	&	1	&	6	&	Personal consumption expenditures: Durable goods: Furnishings and durable equipment	&	FRED-QD	\\
96	&	DREQRG3Q086SBEA	&	1	&	6	&	Personal consumption expenditures: Durable goods: Recreational goods and vehicles	&	FRED-QD	\\
97	&	DODGRG3Q086SBEA	&	1	&	6	&	Personal consumption expenditures: Durable goods: Other durable goods	&	FRED-QD	\\
98	&	DFXARG3Q086SBEA	&	1	&	6	&	Personal consumption expenditures: Nondurable goods: Food and beverages	&	FRED-QD	\\
99	&	DCLORG3Q086SBEA	&	1	&	6	&	Personal consumption expenditures: Nondurable goods: Clothing and footwear	&	FRED-QD	\\
100	&	DGOERG3Q086SBEA	&	1	&	6	&	Personal consumption expenditures: Nondurable goods: Gasoline and other energy goods	&	FRED-QD	\\
101	&	DONGRG3Q086SBEA	&	1	&	6	&	Personal consumption expenditures: Nondurable goods: Other nondurable goods	&	FRED-QD	\\
102	&	DHUTRG3Q086SBEA	&	1	&	6	&	Personal consumption expenditures: Services: Housing and utilities	&	FRED-QD	\\
103	&	DHLCRG3Q086SBEA	&	1	&	6	&	Personal consumption expenditures: Services: Health care	&	FRED-QD	\\
104	&	DTRSRG3Q086SBEA	&	1	&	6	&	Personal consumption expenditures: Transportation services	&	FRED-QD	\\
105	&	DRCARG3Q086SBEA	&	1	&	6	&	Personal consumption expenditures: Recreation services	&	FRED-QD	\\
106	&	DFSARG3Q086SBEA	&	1	&	6	&	Personal consumption expenditures: Services: Food services and accommodations	&	FRED-QD	\\
107	&	DIFSRG3Q086SBEA	&	1	&	6	&	Personal consumption expenditures: Financial services and insurance	&	FRED-QD	\\
108	&	DOTSRG3Q086SBEA	&	1	&	6	&	Personal consumption expenditures: Other services	&	FRED-QD	\\
109	&	CPIAUCSL	&	0	&	6	&	Consumer Price Index for All Urban Consumers: All Items	&	FRED-QD	\\
110	&	CPILFESL	&	0	&	6	&	Consumer Price Index for All Urban Consumers: All Items Less Food \& Energy	&	FRED-QD	\\
111	&	WPSFD49207	&	0	&	6	&	Producer Price Index by Commodity for Final Demand: Finished Goods	&	FRED-QD	\\
112	&	PPIACO	&	0	&	6	&	Producer Price Index for All Commodities	&	FRED-QD	\\
113	&	WPSFD49502	&	1	&	6	&	Producer Price Index by Commodity for Finished Consumer Goods	&	FRED-QD	\\
114	&	WPSFD4111	&	1	&	6	&	Producer Price Index by Commodity for Finished Consumer Foods	&	FRED-QD	\\
115	&	PPIIDC	&	1	&	6	&	Producer Price Index by Commodity Industrial Commodities	&	FRED-QD	\\
116	&	WPSID61	&	1	&	6	&	Producer Price Index by Commodity Intermediate Materials: Supplies \& Components	&	FRED-QD	\\
117	&	WPU0561	&	1	&	5	&	Producer Price Index by Commodity for Fuels and Related Products and Power: Crude Petroleum	&	FRED-QD	\\
118	&	OILPRICEx	&	0	&	5	&	Real Crude Oil Prices: West Texas Intermediate (WTI) - Cushing, Oklahoma	&	FRED-QD	\\
119	&	CES2000000008x	&	0	&	5	&	Real Average Hourly Earnings of Production and Nonsupervisory Employees: Construction	&	FRED-QD	\\
120	&	CES3000000008x	&	0	&	5	&	Real Average Hourly Earnings of Production and Nonsupervisory Employees: Manufacturing	&	FRED-QD	\\
121	&	COMPRNFB	&	1	&	5	&	Manufacturing Sector: Real Compensation Per Hour	&	FRED-QD	\\
122	&	RCPHBS	&	1	&	5	&	Business Sector: Real Compensation Per Hour	&	FRED-QD	\\
123	&	OPHNFB	&	1	&	5	&	Nonfarm Business Sector: Real Output Per Hour of All Persons	&	FRED-QD	\\
124	&	OPHPBS	&	0	&	5	&	Business Sector: Real Output Per Hour of All Persons	&	FRED-QD	\\
125	&	ULCBS	&	0	&	5	&	Business Sector: Unit Labor Cost	&	FRED-QD	\\
126	&	ULCNFB	&	1	&	5	&	Nonfarm Business Sector: Unit Labor Cost	&	FRED-QD	\\
127	&	UNLPNBS	&	1	&	5	&	Nonfarm Business Sector: Unit Nonlabor Payments	&	FRED-QD	\\
128	&	FEDFUNDS	&	1	&	2	&	Effective Federal Funds Rate	&	FRED-QD	\\
129	&	TB3MS	&	1	&	2	&	3-Month Treasury Bill: Secondary Market Rate	&	FRED-QD	\\
130	&	TB6MS	&	0	&	2	&	6-Month Treasury Bill: Secondary Market Rate	&	FRED-QD	\\
131	&	GS1	&	0	&	2	&	1-Year Treasury Constant Maturity Rate	&	FRED-QD	\\
132	&	GS10	&	0	&	2	&	10-Year Treasury Constant Maturity Rate	&	FRED-QD	\\
133	&	AAA	&	0	&	2	&	Moody’s Seasoned Aaa Corporate Bond Yield	&	FRED-QD	\\
134	&	BAA	&	0	&	2	&	Moody’s Seasoned Baa Corporate Bond Yield	&	FRED-QD	\\
135	&	BAA10YM	&	1	&	1	&	Moody’s Seasoned Baa Corporate Bond Yield Relative to Yield on 10-Year Treasury Constant Maturity	&	FRED-QD	\\
136	&	TB6M3Mx	&	1	&	1	&	6-Month Treasury Bill Minus 3-Month Treasury Bill, secondary market	&	FRED-QD	\\
137	&	GS1TB3Mx	&	1	&	1	&	1-Year Treasury Constant Maturity Minus 3-Month Treasury Bill, secondary market	&	FRED-QD	\\
138	&	GS10TB3Mx	&	1	&	1	&	10-Year Treasury Constant Maturity Minus 3-Month Treasury Bill, secondary market	&	FRED-QD	\\
139	&	CPF3MTB3Mx	&	1	&	1	&	3-Month Commercial Paper Minus 3-Month Treasury Bill, secondary market	&	FRED-QD	\\
140	&	AMBSLREAL	&	1	&	5	&	St. Louis Adjusted Monetary Base	&	FRED-QD	\\
141	&	M1REAL	&	1	&	5	&	Real M1 Money Stock	&	FRED-QD	\\
142	&	M2REAL	&	1	&	5	&	Real M2 Money Stock	&	FRED-QD	\\
143	&	MZMREAL	&	1	&	5	&	Real MZM Money Stock	&	FRED-QD	\\
144	&	BUSLOANSx	&	1	&	5	&	Real Commercial and Industrial Loans, All Commercial Banks	&	FRED-QD	\\
145	&	CONSUMERx	&	1	&	5	&	Consumer Loans at All Commercial Banks	&	FRED-QD	\\
146	&	NONREVSLx	&	1	&	5	&	Total Real Nonrevolving Credit Owned and Securitized, Outstanding	&	FRED-QD	\\
147	&	REALLNx	&	1	&	5	&	Real Real-Estate Loans, All Commercial Banks	&	FRED-QD	\\
148	&	TOTALSLx	&	1	&	5	&	Total Consumer Credit Outstanding	&	FRED-QD	\\
149	&	TABSHNOx	&	1	&	5	&	Real Total Assets of Households and Nonprofit Organizations	&	FRED-QD	\\
150	&	TLBSHNOx	&	1	&	5	&	Real Total Liabilities of Households and Nonprofit Organizations	&	FRED-QD	\\
151	&	LIABPIx	&	0	&	5	&	Liabilities of Households and Nonprofit Organizations Relative to Personal Disposable Income	&	FRED-QD	\\
152	&	TNWBSHNOx	&	1	&	5	&	Real Net Worth of Households and Nonprofit Organizations	&	FRED-QD	\\
153	&	NWPIx	&	0	&	1	&	Net Worth of Households and Nonprofit Organizations Relative to Disposable Personal Income	&	FRED-QD	\\
154	&	TARESAx	&	1	&	5	&	Real Assets of Households and Nonprofit Organizations excluding Real Estate Assets	&	FRED-QD	\\
155	&	HNOREMQ027Sx	&	1	&	5	&	Real Real-Estate Assets of Households and Nonprofit Organizations	&	FRED-QD	\\
156	&	TFAABSHNOx	&	1	&	5	&	Real Total Financial Assets of Households and Nonprofit Organizations	&	FRED-QD	\\
157	&	TWEXMMTH	&	1	&	5	&	Trade Weighted U.S. Dollar Index: Major Currencies, Goods	&	FRED-QD	\\
158	&	EXSZUSx	&	1	&	5	&	Switzerland / U.S. Foreign Exchange Rate	&	FRED-QD	\\
159	&	EXJPUSx	&	1	&	5	&	Japan / U.S. Foreign Exchange Rate	&	FRED-QD	\\
160	&	EXUSUKx	&	1	&	5	&	U.S. / U.K. Foreign Exchange Rate	&	FRED-QD	\\
161	&	EXCAUSx	&	1	&	5	&	Canada / U.S. Foreign Exchange Rate	&	FRED-QD	\\
162	&	UMCSENTx	&	0	&	1	&	University of Michigan: Consumer Sentiment	&	FRED-QD	\\
163	&	PAGO	&	1	&	1	&	Current Financial Situation Compared with a Year Ago	&	UofMich	\\
164	&	PEXP	&	1	&	1	&	Expected Change in Financial Situation in a Year	&	UofMich	\\
165	&	NEWS	&	1	&	1	&	News Heard of Recent Changes in Business Conditions	&	UofMich	\\
166	&	BAGO	&	1	&	1	&	Current Business Conditions Compared with a Year Ago	&	UofMich	\\
167	&	BEXP	&	1	&	1	&	Expected Change in Business Conditions in a Year	&	UofMich	\\
168	&	BUS12	&	1	&	1	&	Business Conditions Expected During the Next Year	&	UofMich	\\
169	&	BUS5	&	1	&	1	&	Business Conditions Expected During the Next 5 Years	&	UofMich	\\
170	&	INFEXP	&	1	&	1	&	Expected Change in Prices During the Next Year	&	UofMich	\\
171	&	DUR	&	1	&	1	&	Buying Conditions for Large Household Durables	&	UofMich	\\
172	&	VEH	&	1	&	1	&	Buying Conditions for Vehicles	&	UofMich	\\
173	&	HOM	&	1	&	1	&	Buying Conditions for Houses	&	UofMich	\\
174	&	USASACRQISMEI	&	1	&	1	&	Passenger Car Registrations in United States	&	FRED-QD	\\
175	&	USALOLITONOSTSAM	&	1	&	1	&	Leading indicators: CLI: Normalised for the United States	&	FRED-QD	\\
176	&	BSCICP03USM665S	&	1	&	1	&	Composite Indicators: OECD Indicator for the United States	&	FRED-QD	\\
177	&	B020RE1Q156NBEA	&	0	&	2	&	Shares of gross domestic product: Exports of goods and services	&	FRED-QD	\\
178	&	B021RE1Q156NBEA	&	0	&	2	&	Shares of gross domestic product: Imports of goods and services	&	FRED-QD	\\
179	&	IPMANSICS	&	0	&	5	&	Industrial Production: Manufacturing (SIC)	&	FRED-QD	\\
180	&	IPB51222S	&	0	&	5	&	Industrial Production: Residential Utilities	&	FRED-QD	\\
181	&	IPFUELS	&	0	&	5	&	Industrial Production: Fuels	&	FRED-QD	\\
182	&	UEMPMEAN	&	1	&	2	&	Duration of Unemployment	&	FRED-QD	\\
183	&	CES0600000007	&	1	&	2	&	Average Weekly Hours of Production and Nonsupervisory Employees: Goods-Producing	&	FRED-QD	\\
184	&	TOTRESNS	&	0	&	6	&	Total Reserves of Depository Institutions	&	FRED-QD	\\
185	&	NONBORRES	&	0	&	7	&	Reserves of Depository Institutions, Nonborrowed	&	FRED-QD	\\
186	&	GS5	&	0	&	2	&	5-Year Treasury Constant Maturity Rate	&	FRED-QD	\\
187	&	TB3SMFFM	&	1	&	1	&	3-Month Treasury Constant Maturity Minus Federal Funds Rate	&	FRED-QD	\\
188	&	T5YFFM	&	1	&	1	&	5-Year Treasury Constant Maturity Minus Federal Funds Rate	&	FRED-QD	\\
189	&	AAAFFM	&	1	&	1	&	Moody’s Seasoned Aaa Corporate Bond Minus Federal Funds Rate	&	FRED-QD	\\
190	&	WPSID62	&	1	&	6	&	Producer Price Index: Crude Materials for Further Processing	&	FRED-QD	\\
191	&	PPICMM	&	0	&	6	&	Producer Price Index: Commodities: Metals and metal products: Primary nonferrous metals	&	FRED-QD	\\
192	&	CPIAPPSL	&	0	&	6	&	Producer Price Index: Commodities: Metals and metal products: Primary nonferrous metals	&	FRED-QD	\\
193	&	CPITRNSL	&	1	&	6	&	Consumer Price Index for All Urban Consumers: Transportation	&	FRED-QD	\\
194	&	CPIMEDSL	&	1	&	6	&	Consumer Price Index for All Urban Consumers: Medical Care	&	FRED-QD	\\
195	&	CUSR0000SAC	&	1	&	6	&	Consumer Price Index for All Urban Consumers: Commodities	&	FRED-QD	\\
196	&	CUSR0000SAD	&	1	&	6	&	Consumer Price Index for All Urban Consumers: Durables	&	FRED-QD	\\
197	&	CUSR0000SAS	&	1	&	6	&	Consumer Price Index for All Urban Consumers: Services	&	FRED-QD	\\
198	&	CPIULFSL	&	0	&	6	&	Consumer Price Index for All Urban Consumers: All Items Less Food	&	FRED-QD	\\
199	&	CUSR0000SA0L2	&	0	&	6	&	Consumer Price Index for All Urban Consumers: All items less shelter	&	FRED-QD	\\
200	&	CUSR0000SA0L5	&	0	&	6	&	Consumer Price Index for All Urban Consumers: All items less medical care	&	FRED-QD	\\
201	&	CES0600000008	&	0	&	6	&	Average Hourly Earnings of Production and Nonsupervisory Employees: Goods-Producing	&	FRED-QD	\\
202	&	DTCOLNVHFNM	&	0	&	6	&	Consumer Motor Vehicle Loans Outstanding Owned by Finance Companies	&	FRED-QD	\\
203	&	DTCTHFNM	&	0	&	6	&	Total Consumer Loans and Leases Outstanding Owned and Securitized by Finance Companies	&	FRED-QD	\\
204	&	INVEST	&	1	&	6	&	Securities in Bank Credit at All Commercial Banks	&	FRED-QD	\\
205	&	HWIURATIOx	&	1	&	2	&	Ratio of Help Wanted/No. Unemployed	&	FRED-QD	\\
206	&	CLAIMSx	&	1	&	5	&	Initial Claims	&	FRED-QD	\\
207	&	BUSINVx	&	1	&	5	&	Total Business Inventories	&	FRED-QD	\\
208	&	ISRATIOx	&	1	&	2	&	Total Business: Inventories to Sales Ratio	&	FRED-QD	\\
209	&	CONSPIx	&	0	&	2	&	Nonrevolving consumer credit to Personal Income	&	FRED-QD	\\
210	&	CP3M	&	0	&	2	&	3-Month AA Financial Commercial Paper Rate	&	FRED-QD	\\
211	&	COMPAPFF	&	0	&	1	&	3-Month Commercial Paper Minus Federal Funds Rate	&	FRED-QD	\\
212	&	PERMITNE	&	0	&	5	&	New Private Housing Units Authorized by Building Permits in the Northeast Census Region	&	FRED-QD	\\
213	&	PERMITMW	&	0	&	5	&	New Private Housing Units Authorized by Building Permits in the Midwest Census Region	&	FRED-QD	\\
214	&	PERMITS	&	0	&	5	&	New Private Housing Units Authorized by Building Permits in the South Census Region	&	FRED-QD	\\
215	&	PERMITW	&	0	&	5	&	New Private Housing Units Authorized by Building Permits in the West Census Region	&	FRED-QD	\\
216	&	NIKKEI225	&	0	&	5	&	Nikkei Stock Average	&	FRED-QD	\\
217	&	TLBSNNCBx	&	0	&	5	&	Real Nonfinancial Corporate Business Sector Liabilities	&	FRED-QD	\\
218	&	TLBSNNCBBDIx	&	0	&	1	&	Nonfinancial Corporate Business Sector Liabilities to Disposable Business Income	&	FRED-QD	\\
219	&	TTAABSNNCBx	&	0	&	5	&	Real Nonfinancial Corporate Business Sector Assets	&	FRED-QD	\\
220	&	TNWMVBSNNCBx	&	0	&	5	&	Real Nonfinancial Corporate Business Sector Net Worth	&	FRED-QD	\\
221	&	TNWMVBSNNCBBDIx	&	0	&	2	&	Nonfinancial Corporate Business Sector Net Worth to Disposable Business Income	&	FRED-QD	\\
222	&	TLBSNNBx	&	0	&	5	&	Real Nonfinancial Noncorporate Business Sector Liabilities	&	FRED-QD	\\
223	&	TLBSNNBBDIx	&	0	&	1	&	Nonfinancial Noncorporate Business Sector Liabilities to Disposable Business Income	&	FRED-QD	\\
224	&	TABSNNBx	&	0	&	5	&	Real Nonfinancial Noncorporate Business Sector Assets	&	FRED-QD	\\
225	&	TNWBSNNBx	&	0	&	5	&	Real Nonfinancial Noncorporate Business Sector Net Worth	&	FRED-QD	\\
226	&	TNWBSNNBBDIx	&	0	&	2	&	Nonfinancial Noncorporate Business Sector Net Worth to Disposable Business Income	&	FRED-QD	\\
227	&	CNCFx	&	0	&	5	&	Real Disposable Business Income, Billions of 2009 Dollars	&	FRED-QD	\\
228	&	S\&P 500	&	1	&	5	&	S\&P’s Common Stock Price Index: Composite	&	FRED-QD	\\
229	&	S\&P: indust	&	0	&	5	&	S\&P’s Common Stock Price Index: Industrials	&	FRED-QD	\\
230	&	S\&P div yield	&	0	&	2	&	S\&P’s Composite Common Stock: Dividend Yield	&	FRED-QD	\\
231	&	S\&P PE ratio	&	0	&	5	&	S\&P’s Composite Common Stock: Price-Earnings Ratio	&	FRED-QD	\\
232	&	d/p	&	1	&	2	&	Dividend Price Ratio	&	GW2008	\\
233	&	d/y	&	1	&	2	&	Dividend Yield	&	GW2008	\\
234	&	e/p	&	1	&	2	&	Earnings Price Ratio	&	GW2008	\\
235	&	d/e	&	1	&	2	&	Dividend Payout Ratio	&	GW2008	\\
236	&	b/m	&	1	&	2	&	Book-to-Market Ratio	&	GW2008	\\
237	&	svar	&	1	&	1	&	Stock Market Variance	&	GW2008	\\
238	&	ntis	&	1	&	1	&	Net Equity Expansion	&	GW2008	\\
239	&	lty	&	1	&	1	&	Long Term Yield	&	GW2008	\\
240	&	dfy	&	1	&	1	&	Default Yield Spread	&	GW2008	\\
241	&	dfr	&	1	&	1	&	Default Return Spread	&	GW2008	\\
242	&	Mkt-RF	&	1	&	1	&	Market Excess Return (based on NYSE)	&	K. French	\\
243	&	SMB	&	1	&	1	&	Small Minus Big, Sorted on Size	&	K. French	\\
244	&	HML	&	1	&	1	&	High Minus Low, Sorted on Book-to-Market	&	K. French	\\
245	&	Agric	&	1	&	1	&	Agric Industry Portfolio	&	K. French	\\
246	&	Food 	&	1	&	1	&	Food Industry Portfolio	&	K. French	\\
247	&	Beer 	&	1	&	1	&	Beer Industry Portfolio	&	K. French	\\
248	&	Smoke	&	1	&	1	&	Smoke Industry Portfolio	&	K. French	\\
249	&	Toys 	&	1	&	1	&	Toys Industry Portfolio	&	K. French	\\
250	&	Fun  	&	1	&	1	&	Fun Industry Portfolio	&	K. French	\\
251	&	Books	&	1	&	1	&	Books Industry Portfolio	&	K. French	\\
252	&	Hshld	&	1	&	1	&	Hshld Industry Portfolio	&	K. French	\\
253	&	Clths	&	1	&	1	&	Clths Industry Portfolio	&	K. French	\\
254	&	MedEq	&	1	&	1	&	MedEq Industry Portfolio	&	K. French	\\
255	&	Drugs	&	1	&	1	&	Drugs Industry Portfolio	&	K. French	\\
256	&	Chems	&	1	&	1	&	Chems Industry Portfolio	&	K. French	\\
257	&	Rubbr	&	1	&	1	&	Rubbr Industry Portfolio	&	K. French	\\
258	&	Txtls	&	1	&	1	&	Txtls Industry Portfolio	&	K. French	\\
259	&	BldMt	&	1	&	1	&	BldMt Industry Portfolio	&	K. French	\\
260	&	Cnstr	&	1	&	1	&	Cnstr Industry Portfolio	&	K. French	\\
261	&	Steel	&	1	&	1	&	Steel Industry Portfolio	&	K. French	\\
262	&	Mach 	&	1	&	1	&	Mach Industry Portfolio	&	K. French	\\
263	&	ElcEq	&	1	&	1	&	ElcEq Industry Portfolio	&	K. French	\\
264	&	Autos	&	1	&	1	&	Autos Industry Portfolio	&	K. French	\\
265	&	Aero 	&	1	&	1	&	Aero Industry Portfolio	&	K. French	\\
266	&	Ships	&	1	&	1	&	Ships Industry Portfolio	&	K. French	\\
267	&	Mines	&	1	&	1	&	Mines Industry Portfolio	&	K. French	\\
268	&	Coal 	&	1	&	1	&	Coal Industry Portfolio	&	K. French	\\
269	&	Oil  	&	1	&	1	&	Oil Industry Portfolio	&	K. French	\\
270	&	Util 	&	1	&	1	&	Util Industry Portfolio	&	K. French	\\
271	&	Telcm	&	1	&	1	&	Telcm Industry Portfolio	&	K. French	\\
272	&	PerSv	&	1	&	1	&	PerSv Industry Portfolio	&	K. French	\\
273	&	BusSv	&	1	&	1	&	BusSv Industry Portfolio	&	K. French	\\
274	&	Hardw	&	1	&	1	&	Hardw Industry Portfolio	&	K. French	\\
275	&	Chips	&	1	&	1	&	Chips Industry Portfolio	&	K. French	\\
276	&	LabEq	&	1	&	1	&	LabEq Industry Portfolio	&	K. French	\\
277	&	Paper	&	1	&	1	&	Paper Industry Portfolio	&	K. French	\\
278	&	Boxes	&	1	&	1	&	Boxes Industry Portfolio	&	K. French	\\
279	&	Trans	&	1	&	1	&	Trans Industry Portfolio	&	K. French	\\
280	&	Whlsl	&	1	&	1	&	Whlsl Industry Portfolio	&	K. French	\\
281	&	Rtail	&	1	&	1	&	Rtail Industry Portfolio	&	K. French	\\
282	&	Meals	&	1	&	1	&	Meals Industry Portfolio	&	K. French	\\
283	&	Banks	&	1	&	1	&	Banks Industry Portfolio	&	K. French	\\
284	&	Insur	&	1	&	1	&	Insur Industry Portfolio	&	K. French	\\
285	&	RlEst	&	1	&	1	&	RlEst Industry Portfolio	&	K. French	\\
286	&	Fin  	&	1	&	1	&	Fin Industry Portfolio	&	K. French	\\
287	&	Other	&	1	&	1	&	Other Industry Portfolio	&	K. French	\\
288	&	ME1 BM2	&	1	&	1	&	(1, 2) portfolio sorted on (size, book-to-market)	&	K. French	\\
289	&	ME1 BM3	&	1	&	1	&	(1, 3) portfolio sorted on (size, book-to-market)	&	K. French	\\
290	&	ME1 BM4	&	1	&	1	&	(1, 4) portfolio sorted on (size, book-to-market)	&	K. French	\\
291	&	ME1 BM5	&	1	&	1	&	(1, 5) portfolio sorted on (size, book-to-market)	&	K. French	\\
292	&	ME1 BM6	&	1	&	1	&	(1, 6) portfolio sorted on (size, book-to-market)	&	K. French	\\
293	&	ME1 BM7	&	1	&	1	&	(1, 7) portfolio sorted on (size, book-to-market)	&	K. French	\\
294	&	ME1 BM8	&	1	&	1	&	(1, 8) portfolio sorted on (size, book-to-market)	&	K. French	\\
295	&	ME1 BM9	&	1	&	1	&	(1, 9) portfolio sorted on (size, book-to-market)	&	K. French	\\
296	&	ME1 BM10	&	1	&	1	&	(1, 10) portfolio sorted on (size, book-to-market)	&	K. French	\\
297	&	ME2 BM1	&	1	&	1	&	(2, 1) portfolio sorted on (size, book-to-market)	&	K. French	\\
298	&	ME2 BM2	&	1	&	1	&	(2, 2) portfolio sorted on (size, book-to-market)	&	K. French	\\
299	&	ME2 BM3	&	1	&	1	&	(2, 3) portfolio sorted on (size, book-to-market)	&	K. French	\\
300	&	ME2 BM4	&	1	&	1	&	(2, 4) portfolio sorted on (size, book-to-market)	&	K. French	\\
301	&	ME2 BM5	&	1	&	1	&	(2, 5) portfolio sorted on (size, book-to-market)	&	K. French	\\
302	&	ME2 BM6	&	1	&	1	&	(2, 6) portfolio sorted on (size, book-to-market)	&	K. French	\\
303	&	ME2 BM7	&	1	&	1	&	(2, 7) portfolio sorted on (size, book-to-market)	&	K. French	\\
304	&	ME2 BM8	&	1	&	1	&	(2, 8) portfolio sorted on (size, book-to-market)	&	K. French	\\
305	&	ME2 BM9	&	1	&	1	&	(2, 9) portfolio sorted on (size, book-to-market)	&	K. French	\\
306	&	ME2 BM10	&	1	&	1	&	(2, 10) portfolio sorted on (size, book-to-market)	&	K. French	\\
307	&	ME3 BM1	&	1	&	1	&	(3, 1) portfolio sorted on (size, book-to-market)	&	K. French	\\
308	&	ME3 BM2	&	1	&	1	&	(3, 2) portfolio sorted on (size, book-to-market)	&	K. French	\\
309	&	ME3 BM3	&	1	&	1	&	(3, 3) portfolio sorted on (size, book-to-market)	&	K. French	\\
310	&	ME3 BM4	&	1	&	1	&	(3, 4) portfolio sorted on (size, book-to-market)	&	K. French	\\
311	&	ME3 BM5	&	1	&	1	&	(3, 5) portfolio sorted on (size, book-to-market)	&	K. French	\\
312	&	ME3 BM6	&	1	&	1	&	(3, 6) portfolio sorted on (size, book-to-market)	&	K. French	\\
313	&	ME3 BM7	&	1	&	1	&	(3, 7) portfolio sorted on (size, book-to-market)	&	K. French	\\
314	&	ME3 BM8	&	1	&	1	&	(3, 8) portfolio sorted on (size, book-to-market)	&	K. French	\\
315	&	ME3 BM9	&	1	&	1	&	(3, 9) portfolio sorted on (size, book-to-market)	&	K. French	\\
316	&	ME3 BM10	&	1	&	1	&	(3, 10) portfolio sorted on (size, book-to-market)	&	K. French	\\
317	&	ME4 BM1	&	1	&	1	&	(4, 1) portfolio sorted on (size, book-to-market)	&	K. French	\\
318	&	ME4 BM2	&	1	&	1	&	(4, 2) portfolio sorted on (size, book-to-market)	&	K. French	\\
319	&	ME4 BM3	&	1	&	1	&	(4, 3) portfolio sorted on (size, book-to-market)	&	K. French	\\
320	&	ME4 BM4	&	1	&	1	&	(4, 4) portfolio sorted on (size, book-to-market)	&	K. French	\\
321	&	ME4 BM5	&	1	&	1	&	(4, 5) portfolio sorted on (size, book-to-market)	&	K. French	\\
322	&	ME4 BM6	&	1	&	1	&	(4, 6) portfolio sorted on (size, book-to-market)	&	K. French	\\
323	&	ME4 BM7	&	1	&	1	&	(4, 7) portfolio sorted on (size, book-to-market)	&	K. French	\\
324	&	ME4 BM8	&	1	&	1	&	(4, 8) portfolio sorted on (size, book-to-market)	&	K. French	\\
325	&	ME4 BM9	&	1	&	1	&	(4, 9) portfolio sorted on (size, book-to-market)	&	K. French	\\
326	&	ME4 BM10	&	1	&	1	&	(4, 10) portfolio sorted on (size, book-to-market)	&	K. French	\\
327	&	ME5 BM1	&	1	&	1	&	(5, 1) portfolio sorted on (size, book-to-market)	&	K. French	\\
328	&	ME5 BM2	&	1	&	1	&	(5, 2) portfolio sorted on (size, book-to-market)	&	K. French	\\
329	&	ME5 BM3	&	1	&	1	&	(5, 3) portfolio sorted on (size, book-to-market)	&	K. French	\\
330	&	ME5 BM4	&	1	&	1	&	(5, 4) portfolio sorted on (size, book-to-market)	&	K. French	\\
331	&	ME5 BM5	&	1	&	1	&	(5, 5) portfolio sorted on (size, book-to-market)	&	K. French	\\
332	&	ME5 BM6	&	1	&	1	&	(5, 6) portfolio sorted on (size, book-to-market)	&	K. French	\\
333	&	ME5 BM7	&	1	&	1	&	(5, 7) portfolio sorted on (size, book-to-market)	&	K. French	\\
334	&	ME5 BM8	&	1	&	1	&	(5, 8) portfolio sorted on (size, book-to-market)	&	K. French	\\
335	&	ME5 BM9	&	1	&	1	&	(5, 9) portfolio sorted on (size, book-to-market)	&	K. French	\\
336	&	ME5 BM10	&	1	&	1	&	(5, 10) portfolio sorted on (size, book-to-market)	&	K. French	\\
337	&	ME6 BM1	&	1	&	1	&	(6, 1) portfolio sorted on (size, book-to-market)	&	K. French	\\
338	&	ME6 BM2	&	1	&	1	&	(6, 2) portfolio sorted on (size, book-to-market)	&	K. French	\\
339	&	ME6 BM3	&	1	&	1	&	(6, 3) portfolio sorted on (size, book-to-market)	&	K. French	\\
340	&	ME6 BM4	&	1	&	1	&	(6, 4) portfolio sorted on (size, book-to-market)	&	K. French	\\
341	&	ME6 BM5	&	1	&	1	&	(6, 5) portfolio sorted on (size, book-to-market)	&	K. French	\\
342	&	ME6 BM6	&	1	&	1	&	(6, 6) portfolio sorted on (size, book-to-market)	&	K. French	\\
343	&	ME6 BM7	&	1	&	1	&	(6, 7) portfolio sorted on (size, book-to-market)	&	K. French	\\
344	&	ME6 BM8	&	1	&	1	&	(6, 8) portfolio sorted on (size, book-to-market)	&	K. French	\\
345	&	ME6 BM9	&	1	&	1	&	(6, 9) portfolio sorted on (size, book-to-market)	&	K. French	\\
346	&	ME6 BM10	&	1	&	1	&	(6, 10) portfolio sorted on (size, book-to-market)	&	K. French	\\
347	&	ME7 BM1	&	1	&	1	&	(7, 1) portfolio sorted on (size, book-to-market)	&	K. French	\\
348	&	ME7 BM2	&	1	&	1	&	(7, 2) portfolio sorted on (size, book-to-market)	&	K. French	\\
349	&	ME7 BM3	&	1	&	1	&	(7, 3) portfolio sorted on (size, book-to-market)	&	K. French	\\
350	&	ME7 BM4	&	1	&	1	&	(7, 4) portfolio sorted on (size, book-to-market)	&	K. French	\\
351	&	ME7 BM5	&	1	&	1	&	(7, 5) portfolio sorted on (size, book-to-market)	&	K. French	\\
352	&	ME7 BM6	&	1	&	1	&	(7, 6) portfolio sorted on (size, book-to-market)	&	K. French	\\
353	&	ME7 BM7	&	1	&	1	&	(7, 7) portfolio sorted on (size, book-to-market)	&	K. French	\\
354	&	ME7 BM8	&	1	&	1	&	(7, 8) portfolio sorted on (size, book-to-market)	&	K. French	\\
355	&	ME7 BM9	&	1	&	1	&	(7, 9) portfolio sorted on (size, book-to-market)	&	K. French	\\
356	&	ME7 BM10	&	1	&	1	&	(7, 10) portfolio sorted on (size, book-to-market)	&	K. French	\\
357	&	ME8 BM1	&	1	&	1	&	(8, 1) portfolio sorted on (size, book-to-market)	&	K. French	\\
358	&	ME8 BM2	&	1	&	1	&	(8, 2) portfolio sorted on (size, book-to-market)	&	K. French	\\
359	&	ME8 BM3	&	1	&	1	&	(8, 3) portfolio sorted on (size, book-to-market)	&	K. French	\\
360	&	ME8 BM4	&	1	&	1	&	(8, 4) portfolio sorted on (size, book-to-market)	&	K. French	\\
361	&	ME8 BM5	&	1	&	1	&	(8, 5) portfolio sorted on (size, book-to-market)	&	K. French	\\
362	&	ME8 BM6	&	1	&	1	&	(8, 6) portfolio sorted on (size, book-to-market)	&	K. French	\\
363	&	ME8 BM7	&	1	&	1	&	(8, 7) portfolio sorted on (size, book-to-market)	&	K. French	\\
364	&	ME8 BM8	&	1	&	1	&	(8, 8) portfolio sorted on (size, book-to-market)	&	K. French	\\
365	&	ME8 BM9	&	1	&	1	&	(8, 9) portfolio sorted on (size, book-to-market)	&	K. French	\\
366	&	ME8 BM10	&	1	&	1	&	(8, 10) portfolio sorted on (size, book-to-market)	&	K. French	\\
367	&	ME9 BM1	&	1	&	1	&	(9, 1) portfolio sorted on (size, book-to-market)	&	K. French	\\
368	&	ME9 BM2	&	1	&	1	&	(9, 2) portfolio sorted on (size, book-to-market)	&	K. French	\\
369	&	ME9 BM3	&	1	&	1	&	(9, 3) portfolio sorted on (size, book-to-market)	&	K. French	\\
370	&	ME9 BM4	&	1	&	1	&	(9, 4) portfolio sorted on (size, book-to-market)	&	K. French	\\
371	&	ME9 BM5	&	1	&	1	&	(9, 5) portfolio sorted on (size, book-to-market)	&	K. French	\\
372	&	ME9 BM6	&	1	&	1	&	(9, 6) portfolio sorted on (size, book-to-market)	&	K. French	\\
373	&	ME9 BM7	&	1	&	1	&	(9, 7) portfolio sorted on (size, book-to-market)	&	K. French	\\
374	&	ME9 BM8	&	1	&	1	&	(9, 8) portfolio sorted on (size, book-to-market)	&	K. French	\\
375	&	ME9 BM10	&	1	&	1	&	(9, 10) portfolio sorted on (size, book-to-market)	&	K. French	\\
376	&	ME10 BM1	&	1	&	1	&	(10, 1) portfolio sorted on (size, book-to-market)	&	K. French	\\
377	&	ME10 BM2	&	1	&	1	&	(10, 2) portfolio sorted on (size, book-to-market)	&	K. French	\\
378	&	ME10 BM3	&	1	&	1	&	(10, 3) portfolio sorted on (size, book-to-market)	&	K. French	\\
379	&	ME10 BM4	&	1	&	1	&	(10, 4) portfolio sorted on (size, book-to-market)	&	K. French	\\
380	&	ME10 BM5	&	1	&	1	&	(10, 5) portfolio sorted on (size, book-to-market)	&	K. French	\\
381	&	ME10 BM6	&	1	&	1	&	(10, 6) portfolio sorted on (size, book-to-market)	&	K. French	\\
382	&	ME10 BM7	&	1	&	1	&	(10, 7) portfolio sorted on (size, book-to-market)	&	K. French	\\
383	&	Natural gas index	&	1	&	5	&	Commodity Prices, Natural Gas Index	&	World Bank	\\
384	&	Cocoa	&	1	&	5	&	Commodity Prices, Cocoa	&	World Bank	\\
385	&	Coffee, Arabica	&	1	&	5	&	Commodity Prices, Coffee, Arabica	&	World Bank	\\
386	&	Coffee, Robusta	&	1	&	5	&	Commodity Prices, Coffee, Robusta	&	World Bank	\\
387	&	Tea	&	1	&	5	&	Commodity Prices, Tea, avg 3 auctions	&	World Bank	\\
388	&	Tea, Colombo	&	1	&	5	&	Commodity Prices, Tea, Colombo	&	World Bank	\\
389	&	Tea, Kolkata	&	1	&	5	&	Commodity Prices, Tea, Kolkata	&	World Bank	\\
390	&	Tea, Mombasa	&	1	&	5	&	Commodity Prices, Tea, Mombasa	&	World Bank	\\
391	&	Coconut oil	&	1	&	5	&	Commodity Prices, Coconut Oil	&	World Bank	\\
392	&	Groundnut oil	&	1	&	5	&	Commodity Prices, Groundnut Oil	&	World Bank	\\
393	&	Palm oil	&	1	&	5	&	Commodity Prices, Palm Oil	&	World Bank	\\
394	&	Soybeans	&	1	&	5	&	Commodity Prices, Soybeans	&	World Bank	\\
395	&	Soybean oil	&	1	&	5	&	Commodity Prices, Soybean Oil	&	World Bank	\\
396	&	Soybean meal	&	1	&	5	&	Commodity Prices, Soybean Meal	&	World Bank	\\
397	&	Barley	&	1	&	5	&	Commodity Prices, Barley	&	World Bank	\\
398	&	Maize	&	1	&	5	&	Commodity Prices, Maize	&	World Bank	\\
399	&	Sorghum	&	1	&	5	&	Commodity Prices, Sorghum	&	World Bank	\\
400	&	Rice	&	1	&	5	&	Commodity Prices, Rice, Thai 5\% 	&	World Bank	\\
401	&	Wheat	&	1	&	5	&	Commodity Prices, Wheat, US HRW	&	World Bank	\\
402	&	Banana	&	1	&	5	&	Commodity Prices, Banana, US	&	World Bank	\\
403	&	Orange	&	1	&	5	&	Commodity Prices, Orange	&	World Bank	\\
404	&	Beef	&	1	&	5	&	Commodity Prices, Beef	&	World Bank	\\
405	&	Chicken	&	1	&	5	&	Commodity Prices, Meat, Chicken	&	World Bank	\\
406	&	Shrimps	&	1	&	5	&	Commodity Prices, Shrimps, Mexican	&	World Bank	\\
407	&	Sugar	&	1	&	5	&	Commodity Prices, Sugar, World	&	World Bank	\\
408	&	Tobacco	&	1	&	5	&	Commodity Prices, Tobacco, US import u.v.	&	World Bank	\\
409	&	Logs	&	1	&	5	&	Commodity Prices, Logs, Malaysian	&	World Bank	\\
410	&	Sawnwood	&	1	&	5	&	Commodity Prices, Sawnwood, Malaysian	&	World Bank	\\
411	&	Cotton	&	1	&	5	&	Commodity Prices, Cotton, A Index	&	World Bank	\\
412	&	Rubber	&	1	&	5	&	Commodity Prices, Rubber, SGP/MYS	&	World Bank	\\
413	&	Copper	&	1	&	5	&	Commodity Prices, Copper	&	World Bank	\\
414	&	Lead	&	1	&	5	&	Commodity Prices, Lead	&	World Bank	\\
415	&	Tin	&	1	&	5	&	Commodity Prices, Tin	&	World Bank	\\
416	&	Nickel	&	1	&	5	&	Commodity Prices, Nickel	&	World Bank	\\
417	&	Zinc	&	1	&	5	&	Commodity Prices, Zinc	&	World Bank	\\
418	&	Gold	&	1	&	5	&	Commodity Prices, Gold	&	World Bank	\\
419	&	Platinum	&	1	&	5	&	Commodity Prices, Platinum	&	World Bank	\\
420	&	Silver	&	1	&	5	&	Commodity Prices, Silver	&	World Bank	\\
421	&	JPNPROINDQISMEI	&	1	&	5	&	Production of Total Industry in Japan	&	FRED	\\
422	&	LRHUTTTTJPQ156S	&	1	&	5	&	Harmonized Unemployment Rate: Total: All Persons for Japan	&	FRED	\\
423	&	JPNCPIALLQINMEI	&	1	&	5	&	Consumer Price Index of All Items in Japan	&	FRED	\\
424	&	JPNLOLITONOSTSAM	&	1	&	1	&	Leading indicators: CLI: Normalised for Japan	&	FRED	\\
425	&	DEUPROINDQISMEI	&	1	&	5	&	Production of Total Industry in Germany	&	FRED	\\
426	&	OPCNRE01DEQ661N	&	1	&	5	&	Total Cost of Residential Construction for Germany	&	FRED	\\
427	&	IRLTLT01DEQ156N	&	1	&	2	&	Long-Term (10-year) Government Bond Yields for Germany	&	FRED	\\
428	&	DEUCPIALLQINMEI	&	1	&	5	&	Consumer Price Index of All Items in Germany	&	FRED	\\
429	&	SPASTT01DEQ661N	&	1	&	5	&	Total Share Prices for All Shares for Germany	&	FRED	\\
430	&	QDEPAMUSDA	&	1	&	5	&	Total Credit to Private Non-Financial Sector for Germany	&	FRED	\\
431	&	GBRPROINDQISMEI	&	1	&	5	&	Production of Total Industry in the United Kingdom	&	FRED	\\
432	&	IRLTLT01GBQ156N	&	1	&	2	&	Long-Term (10-year) Government Bond Yields for the United Kingdom	&	FRED	\\
433	&	GBRCPIALLQINMEI	&	1	&	5	&	Consumer Price Index of All Items in the United Kingdom	&	FRED	\\
434	&	LMUNRRTTGBQ156S	&	1	&	2	&	Registered Unemployment Rate for the United Kingdom	&	FRED	\\
435	&	SPASTT01GBQ661N	&	1	&	5	&	Total Share Prices for All Shares for the United Kingdom	&	FRED	\\
436	&	GBRGFCFQDSMEI	&	1	&	5	&	Gross Fixed Capital Formation in United Kingdom	&	FRED	\\
437	&	GBRLOLITONOSTSAM	&	1	&	1	&	Leading indicators: CLI: Normalised for the United Kingdom	&	FRED	\\
438	&	CANPROINDQISMEI	&	1	&	5	&	Production of Total Industry in Canada	&	FRED	\\
439	&	WSCNDW01CAQ489S	&	1	&	4	&	Total Dwellings and Residential Buildings by Stage of Construction, Started for Canada	&	FRED	\\
440	&	IRLTLT01CAQ156N	&	1	&	2	&	Long-Term (10-year) Government Bond Yields for Canada	&	FRED	\\
441	&	LRUNTTTTCAQ156S	&	1	&	2	&	Unemployment Rate: Aged 15 and Over: All Persons for Canada	&	FRED	\\
442	&	QCAPAM770A	&	1	&	5	&	Total Credit to Private Non-Financial Sector for Canada	&	FRED	\\
443	&	SPASTT01CAQ661N	&	1	&	5	&	Total Share Prices for All Shares for Canada	&	FRED	\\
444	&	CANLOLITONOSTSAM	&	1	&	1	&	Leading indicators: CLI: Normalised for Canada	&	FRED	\\
\end{longtable}
}
\end{landscape}

\end{appendix}

\end{document}